\documentclass{article}
\usepackage{amsfonts,amsmath,amsthm,latexsym,amscd,graphicx,xypic,psfrag}
\usepackage{epic, eepic}
\usepackage{bbold, mathabx}
\usepackage{setspace}
\usepackage[colorlinks=true]{hyperref}
\hypersetup{urlcolor=blue, citecolor=red}

\def\<#1,#2>{\langle #1,#2 \rangle}

\def\bothID{\rlap{\hbox to.97\wd0{\hss\vrule height.06\ht0 width.82\wd0}}
 \copy0\rlap{\kern-.36\wd0\vrule height1.05\ht0 width.05\ht0}\kern.14\wd0}

\DeclareMathOperator{\diag}{diag}

\theoremstyle{definition}  %
\newtheorem{rem}{Remark}[section]  %
\newtheorem*{defi}{Definition}

\begin{document}

\title{Hydroassets Portfolio Management for Intraday Electricity Trading from
a Discrete Time Stochastic Optimization Perspective}

\author{Simone Farinelli \\
Core Dynamics GmbH\\
Scheuchzerstrasse 43\\
CH-8006 Zurich, Switzerland\\
Email: simone@coredynamics.ch\\
and\\
Luisa Tibiletti \\
Department of Management\\
University of Torino\\
Corso Unione Sovietica 218 bis I-10134 Torino\\
Email: luisa.tibiletti@unito.it }
\maketitle

\maketitle

\maketitle

\begin{abstract}
Hydro storage system optimization is becoming one of the most challenging
tasks in Energy Finance. While currently the state-of-the-art of the commercial
software in the industry implements mainly linear models, we would like to introduce risk
aversion and a generic utility function. At the same time, we aim to develop and implement
a computational efficient algorithm, which is not affected by the curse of dimensionality
 and does not utilize subjective heuristics to prevent it. For the short term power market we propose
 a simultaneous solution for both dispatch and bidding problems.\par
  Following the Blomvall and Lindberg (2002) interior
point model, we set up a stochastic multiperiod optimization procedure by
means of a \textquotedblright bushy\textquotedblright\ recombining tree that
provides fast computational results. Inequality constraints are packed into
the objective function by the logarithmic barrier approach and the utility
function is approximated by its second order Taylor polynomial. The optimal
solution for the original problem is obtained as a diagonal sequence where
the first diagonal dimension is the parameter controlling the logarithmic
penalty and the second one is the parameter for the Newton step in the
construction of the approximated solution. Optimal intraday electricity
trading and water values for hydroassets as shadow prices are computed. The
algorithm is implemented in Mathematica.
\textit{Keywords:} Stochastic multiperiod optimization \and Stochastic market \and
Blomvall and Lindberg interior point model \and Logarithmic barrier approach \and
Energy markets \and Spot and intraday prices
\textit{PACS:}49M15 \and 49M37 \and 90C15 \and 90C30 \and 90C39 \and 90C51
\end{abstract}

\section{Introduction}

The liberalised electricity market poses new challenges to power generating
companies for the electrical grid. A key driver to set up economically
efficient grids is the capacity to store electricity through hydro storage
systems and thereby decouple electricity generation from electricity
consumption. So, the hydro storage system optimization is becoming one of
the most challenging tasks in Energy Finance, as highlighted in \cite{LMW13}
and in \cite{ESS14}. While the current industrial standard for hydro optimization
covers linear models, recently risk aversion optimizations, which are very
common in financial portfolio optimization, have been introduced into the
energy sector, see f.i. \cite{Ab15} and \cite{PKS15}.\par

The aim of this research work is to set up a computational efficiently implementable concave stochastic dynamic
program in order to optimize intraday electricity trading under risk aversion,  and
model at the same time water values for hydro assets. It extends the previous
work of the authors (\cite{FT15}) by presenting the complete algorithm and
constructing numerical examples. Its two main contributions are:
\begin{itemize}
\item The implementation of the optimization algorithm of Blomvall and Lindberg on a lattice guaranteeing computational efficiency. To our knowledge this approach is new and can be utilized for the discretization of virtually any intertemporal portfolio optimization.
\item The introduction of deterministic water values of an hydro infrastructutre as certainty equivalents of optimal stochastic Lagrangian multipliers corresponding to the basin level equations.
\end{itemize}
The optimization of electricity trading under risk aversion is formulated as a stochastic
multiperiod optimization problem in discrete time for a generic utility
function. More exactly, the objective function is the weighted sum of the expected utility
of the wealth generated by the electricity trading during each subinterval. The optimization problem is
subject to equality restrictions, such as the equations for the levels of all basins and to inequality restrictions, such as
the lower and upper bounds for the levels of all basins or the limits for the turbined or pumped water.
For linear restrictions and a generic concave utility function this optimization problem is known to have always a unique solution,
an optimal (stochastic) dynamic dispatch plan. However, in general, an explicit solution cannot be computed directly but can only be approximated
by a sequence of suboptimal dispatch plans. These can be obtained following
the seminal Blomvall and Lindberg's ideas (see \cite{BL00}, \cite{BL02}, \cite{BL02Bis}, \cite{BL03} and \cite{BL03Bis}),
where inequality constraints are packed into the objective
function by means of an additive logarithmic penalty - a technique known as logarithmic barrier approach. The optimization problem with the barrier approximates the original one and can be solved by a Newton's scheme, where the utility function is
approximated by its second order Taylor polynomial. This newly obtained quadratic optimization problem, approximates again the original one, and has an explicit closed formula solution, which depends on two parameters: the first one is the parameter controlling the logarithmic penalty and the second one is the step parameter in Newton's scheme. Finally, the optimal solution for the original
problem is obtained as a diagonal sequence over this two parameters.\par
We provide generic formulae in terms of conditional expectations and thus not depending on the way the underlying stochastic processes are modelled for the original deterministic equivalent formulation as in Blomvall and Lindberg.
In the practical implementation intraday prices and water inflows are discretized in the space dimensions by means of a ``bushy''
recombining tree (meaning by this a $k$-dimensional lattice with $k>>1$), so that we are not worried by
the dimensionality curse nor we have to deal with heuristic arguments
concerning the choice of representative branches in a non recombining ``sparse'' tree,
as Blomvall and Lindberg implicitly have to deal with in their original
work. For a more recent treatment of scenario reduction techniques in stochastic programming we refer to \cite{PP14} and \cite{Roe09}.\par
The obtained algorithm is implemented in Mathematica and applied to optimize
intraday electricity trading and model at the same time stochastic water values for
hydro assets. These are defined as shadow prices, that is the optimal
Lagrangian multipliers associated with the equality restrictions given by the equations for the basin levels.
Deterministic water values are obtained by passing to the certainty equivalents.\par
This paper is structured as follows. Section 2 introduces the set up for
discrete intertemporal expected utility optimization of portfolio subject to
constraints, solved by means of an algorithm developed in Section 3, where Remark \ref{diff} highlights the differences between
Blomvall and Lindberg's work and our proposed approach. Section
4 deals with the implementation of the solution method on a lattice, seen as
recombing tree. This is applied in Section 5 to the intraday electricity
trading to find an optimal strategy and to determine water values of
hydro electric infrastructures to be used for market bids. Section 6 presents
a numerical example. Section 7 concludes.
\subsection{A Short Review of the Literature}
Energy trading methods have been widely studied in the technical literature in the past $20$ years. References \cite{FW03}, \cite{La04} and \cite{Yam04} are some of the few reviews about different algorithms applied to hydro power planning. Some of these
techniques became standard in solving of medium-term hydro power planning problems. The pioneering research of R. Bellman (\cite{Bel57}) introduced and made popular the framework of dynamic programming, which was very soon extended to stochastic dynamic programming to account for the uncertainties of the underlying processes.
With randomly variable inflows and consumption (electricity prices were liberalized only in the $1990$s) hydro power scheduling was therefore used as an application example for stochastic dynamic programming from the beginning. But because of its computational challenging nature the problem was first solved for a single basin configuration only at the end of the $1960$s (see \cite{Yo67}) and was an active field of research during the $1970$s and the early $1980$s as the comprehensive reviews \cite{LSS84} and \cite{Yak82} show. The basic algorithms were extended
to better account for stochasticity, multi reservoirs, hydro thermal systems, reliability
constraints, and improving the model for water inflows. During the $1990$s, thanks to the increase of computing power, approximate dynamic programming and, in particular stochastic dual dynamic programming, was in the spotlight. For the techniques allowing to approximate
some of the problem's elements and reducing the computational time we refer to the description of many of the algorithms in question, which can be found in \cite{Ber05} and \cite{Po11}.\par
Originally, risk aversion was introduced into hydro power
production in order to achieve a certain reliability, which was mainly expressed in terms of constraints for the optimization problem (e.g. \cite{As74}, \cite{Ro77} and \cite{Sn79}). With the liberalization of electricity markets the attention was
focused on profit risk mitigation. In terms of modelling this was achieved first by similar
methods, i.e. by setting target ranges for some variables (e. g. \cite{GGM01}).
In more recent years, following the discussion on coherent risk measures (\cite{ADEH99}) first and time consistency of risk measures (\cite{Ri04}) later, stochastic dynamic programming has considered risk measures in the objective function depending on the control rules and on the underlying stochastic processes. Applications to hydro power production can be found in \cite{De07}, \cite{Ru10}, \cite{CSST12} and \cite{PP14}.\par
We remark that risk aversion optimization can be formulated by choosing the objective function as a trade off between reward and risk, or, by setting the objective function equal to the expected utility for a concave utility function. The latter is the approach followed in this paper, where by means of risk averse stochastic dynamic programming applied to the intraday electricity market, we derive optimal short term dispatch plans and appropriate hydro infrastructure water values for the day ahead market bids. Of course this model can be extended to arbitrary long time horizons, for which the risk aversion plays an even more important role, if the whole dynamics of the hourly priced forward curve and not just the intraday prices are considered.\par
In \cite{GGMP07} a mixed-integer linear program maximizes the expected profit of a hydro chain in the day-ahead market, avoiding unnecessary spillages and considering start-up costs. In \cite{LM10} expected discounted cash flows of rewards are maximized without taking risk aversion into account. But, for computational efficiency, instead of linear programming, an approximated stochastic dynamic programming algorithm is utilized, which consists in a combination of temporal difference learning and least squares policy evaluation. In \cite{FK07} and \cite{FK08} a two stage mixed integer-linear program maximizes a trade off between the expected profit for the one-day operation and a penalty/reward for imbalances in the future production. Being the objective function linear, there is no explicit risk aversion.
While the first stage determines the one-day production plan and involves the bidding process, the second stage evaluates the impact of the one-day production plan on future production. The output is an optimal bid for the day-ahead market in terms of volumes and prices and an optimal dispatch plan. For a similar problem set up \cite{LMW13} efficiently solve a stochastic mixed-integer quadratic program integrating stochastic dynamic programming with ideas of approximate dynamic programming.\par
Recent references giving a thorough overview of producer models for bidding in the auction market with and without a dispatch plan are \cite{FK07} (mixed integer programming), \cite{LSQ11} (mathematical programming, game theory and agent-based models), \cite{BRS14} (simulation, various forms of integer programming, various forms of dynamic programming, equilibrium models, evolutionary algorithms), and \cite{FK10} (stochastic programming models in short term power generation scheduling and bidding). Similar problems in economic dispatch are solved in \cite{ABIPR17} by means of a oblivious routing economic dispatch algorithm.\par
How does our work fit into this model landscape? It has the following characteristics:
\begin{itemize}
\item It is a convex risk averse optimization problem.
\item It is solved for a generic utility function.
\item It utilizes stochastic dynamic programming and the Bellman recursion.
\item It is implemented on fully recombining tree avoiding the curse of dimensionality.
\item It solves the scheduling and the bidding problem simultaneously.
\end{itemize}

\subsection{Overview of the Nomenclature and of the Document Structure}
\vspace{-0.75cm}
{\footnotesize
\begin{flalign*}
&T:\text{ Final time horizon } & (\ref{sec2})\\
&t=0,1,2,3,\dots,T:\text{ Time points } & (\ref{sec2})\\
&(\Omega, \mathcal{A}, (\mathcal{A}_t)_{t=0,\dots T}, P):\text{ Filtered probability space } & (\ref{sec2})\\
&\mathbb{E}_0[\cdot]:\text{ Statistical expectation } & (\ref{sec2})\\
&\mathbb{E}_t[\cdot]:\text{ Statistical conditional expectation at time $t$ } & (\ref{subsec33})\\
&Z_0, Z_1,Z_2,Z_3,\dots, Z_T:\text{ Risk drivers } & (\ref{sec2})\\
&K:\text{ Dimension of risk drivers } & (\ref{sec2})\\
&X_0, X_1,X_2,X_3,\dots, X_T:\text{ External states (or risk factors) } & (\ref{sec2})\\
&N:\text{ Dimension of external states } & (\ref{sec2})\\
&u_0, u_1,u_2,u_3,\dots, u_{T-1}:\text{ Control rules } & (\ref{sec2})\\
&Y_0, Y_1,Y_2,Y_3,\dots, Y_T:\text{ Internal states (functions of external states and control rules) } & (\ref{sec2})\\
&M:\text{ Dimension of internal states } & (\ref{sec2})\\
&U:\text{ Utility function } & (\ref{sec2})\\
&V_t:\text{ Portfolio value at time $t$ } & (\ref{sec2})\\
&\mathcal{C}:\text{ Set of linear equality and inequality constraints } & (\ref{sec2})\\
&\mathcal{C}_{\text{ineq}}:\text{ Set of linear inequality constraints } & (\ref{sec3})\\
&(E_t)_{t}\subset\mathbf{R}^{L\times M},
(F_t)_{t}\subset\mathbf{R}^{L\times N}, (e_t)_{t}\subset\mathbf{R}^{L\times 1}:\text{ Processes utilized to express linear inequality constraints } & (\ref{sec3})\\
&\mathcal{C}_{\text{eq}}:\text{ Set of linear equality constraints } & (\ref{sec3})\\
&(A_t)_{t}\subset\mathbf{R}^{M\times M}, (B_t)_{t}\subset\mathbf{R}^{M\times N}, (b_t)_{t}\subset\mathbf{R}^{M\times 1}:\text{ Processes utilized to express linear equality constraints } & (\ref{sec3})\\
&(\beta_t)_{t=1,\dots T}>0:\text{ Positive deterministic weights } & (\ref{sec2})\\
&\mu: \text{ Trade off parameter between expected utility and penalty function induced by the restrictions } & (\ref{sec3})\\
&1: \text{ Vector of ones  } & (\ref{subsec31})\\
&\Phi: \text{ Lagrange principal function } & (\ref{subsec32})\\
&y_{\ge t}:=(y_s)_{s\ge t}, u_{\ge t}:=(u_s)_{s\ge t}: \text{ Internal states and control rules from time $t$ till the end }&(\ref{subsec33})\\
&h_t: \text{ Quadratic Taylor polynomial of objective function at time $t$ } & (\ref{subsec33})\\
&q_t: \text{ Gradient of $h_t$ with respect to internal states $y_{\ge t}$ } & (\ref{subsec33})\\
&r_t: \text{ Gradient of $h_t$ with respect to control rules $u_{\ge t}$} & (\ref{subsec33})\\
&Q_t, P_t, R_t: \text{ Submatrices of the Hessian of $h_t$ with respect to internal states and control rules } & (\ref{subsec33})\\
&J_t: \text{ Value function at time $t$ for the Bellman recursion of the optimization problem} & (\ref{subsec34})\\
&\overline{q}_t: \text{ Gradient of $h_t$ with respect to internal states $y_t$ } & (\ref{subsec34})\\
&\overline{r}_t: \text{ Gradient of $h_t$ with respect to control rules $u_t$ } & (\ref{subsec34})\\
&\overline{Q}_t, \overline{P}_t, \overline{R}_t: \text{ Submatrices of the Hessian of $h_t$ with respect to internal states $y_t$ and control rules $u_t$ } & (\ref{subsec34})\\
&(W_t)_{t}, (\alpha_t)_{t}, (w_t)_{t}, (\widetilde{a}_t)_t, (\widetilde{r}_t)_t, (\widetilde{R}_t)_t, (\widetilde{q}_t)_,
(\widetilde{Q}_t)_t, (\widetilde{P}_t)_t:\text{  Adapted
processes utilized in the inductive assumption for $(J_t)_t$ } & (\ref{subsec34})\\
&u_t^*: \text{ Optimal control rule } & (\ref{subsec35})\\
&(\alpha_t)_t, (w_t)_t, (W_t)_t: \text{  Adapted
processes utilized in the Riccati equation } & (\ref{subsec35})
\end{flalign*}

\begin{flalign*}
&\mathcal{L}: \text{ Lattice } & (\ref{sec4})\\
&\mathcal{L}_t: \text{ Time $t$ layer of lattice } & (\ref{sec4})\\
&k: \text{ Number of branches for every node  in the lattice } & (\ref{sec4})\\
&n_t(i):\text{ Node in lattice layer at time $t$ } & (\ref{sec4})\\
&\text{Children}(n_t(i)):\text{ Children of node $n_t(i)$ } & (\ref{sec4})\\
&\text{Parents}(n_s(j)):\text{  Parents of node $n_s(j)$ } & (\ref{sec4})\\
&N_t:\text{ Number of nodes in lattice layer at time $t$ } & (\ref{sec4})\\
&\mathcal{N}_T:\text{ Number of nodes in lattice } & (\ref{sec4})\\
&z_t^1,\dots,z_t^{N_t}:\text{ Simulated values for the risk drivers on the lattice layer at time $t$ } & (\ref{sec4})\\
&\epsilon_t:\text{ Contraction factor for $\Delta u_t$ which guarantees feasibility in every Newton step } & (\ref{sec4})\\
&\mathcal{B}(n_t):\text{ Atom associated to the node $n_t$ of the $\sigma$ algebra $\mathcal{A}_t$ for the time $t$ lattice layer } & (\ref{sec4})\\
&S_t:\text{ Spot electricity price } & (\ref{sec5})\\
&\text{GP}_t^\text{Bid},\text{GP}_t^\text{Ask}:\text{ Electricity bid and ask prices in the day ahead market bidding } & (\ref{sec5})\\
&\Xi_t^{\text{Bid}},\Xi_t^{\text{Ask}}:\text{ Electricity bid and ask volumes in the day ahead market bidding } & (\ref{sec5})\\
&\Xi_t^{\text{Spot, Sell}},\Xi_t^{\text{Spot, Buy}}:\text{ Electricity sell and buy volumes in the day ahead market } & (\ref{sec5})\\
&B:\text{ Number of basins } & (\ref{sec5})\\
&\text{gp}_t^{\text{Ask}}:\text{ Stochastic water value } & (\ref{sec5})\\
&F_t: \text{ Forward price } & (\ref{sec5})\\
&\Psi_t: \text{ Energy volume for the forward market } & (\ref{sec5})\\
&\mathbb{E}_0[r]: \text{ Reward measure } & (\ref{sec5})\\
&\mathbb{E}_0[\rho]: \text{ Risk measure } & (\ref{sec5})\\
&w: \text{ Risk aversion } & (\ref{sec5})\\
\end{flalign*}
}

\begin{figure}[!]
  \centering
  \includegraphics[angle=-90, width=10cm]{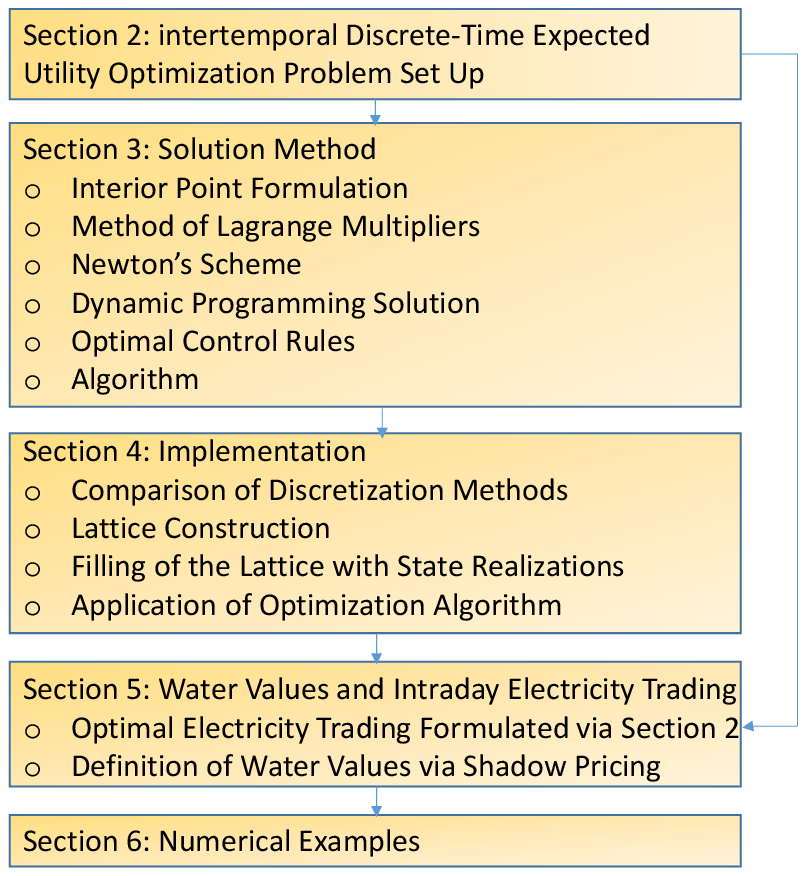}\\
  \caption{Document Structure}\label{Scheme}
\end{figure}

\section{Discrete Multiperiod Portfolio Expected Utility Maximization}\label{sec2}
The purpose of this section is to show how the intertemporal expected
utility framework can be used  to solve optimization problems for a portfolio of financial assets (Example $1$)
or for the power production of an hydro infrastructure (Example $2$). We first introduce the necessary notation for the discrete time setting
given a final time horizon $T$, time points $t=0,1,2,3,\dots,T$ and a filtered probability space $(\Omega, \mathcal{A}, (\mathcal{A}_t)_{t=0,\dots T}, P)$ with statistical expectation $\mathbb{E}_0[\cdot]$:

\begin{itemize}
\item \textbf{Risk drivers:}\\ $Z_0, Z_1,Z_2,Z_3,\dots, Z_T$, where $%
Z_t:\Omega\rightarrow\mathbf{R}^K$ is a $t$-measurable random variable. The
random variables $(Z_t)_t$ are assumed to be i.i.d.

\item \textbf{External states (or risk factors):}\\ $X_0, X_1,X_2,X_3,\dots, X_T$,
where $X_t:\Omega\rightarrow\mathbf{R}^M$ is a $t$-measurable random
variable. We assume that there exists deterministic functions $%
(f_t)_{t=1,\dots T}$ such that $X_t=f_t(X_{t-1},Z_t)$.
\begin{itemize}
 \item \textbf{Ex1:} Asset values for the different asset classes.
 \item \textbf{Ex2:} Hourly electricity price for the intraday market, hourly water inflows for the basins.
\end{itemize}
\item \textbf{Control rules:}\\ $u_0, u_1,u_2,u_3,\dots, u_{T-1}$, where $%
u_t:\Omega\rightarrow\mathbf{R}^N$ is a $t$-measurable random variable.
\begin{itemize}
 \item \textbf{Ex1:} Holdings in the different asset classes.
 \item \textbf{Ex2:} Water processed by the different turbines and pumps of the hydro infrastructure.
\end{itemize}

\item \textbf{Internal states (functions of external states and control rules):}\\
$Y_0, Y_1,Y_2,Y_3,\dots, Y_T$, where $Y_t:\Omega\rightarrow\mathbf{R}^M$ is a
$t$-measurable random variable. We assume that there exists deterministic
functions $(g_t)_{t=1,\dots T}$ such that $Y_t=g_t(Y_{t-1},u_{t-1},X_t)$.

\begin{itemize}
 \item \textbf{Ex1:} Wealth level for the portfolio.
 \item \textbf{Ex2:} Basin levels of the hydro infrastructure.
\end{itemize}

\item \textbf{Utility function:}\\ a concave, monotone increasing differentiable
function $U:\mathbf{R}\hookrightarrow\mathbf{R}$

\item \textbf{Portfolio value:}\\ $V_t=V_t(X_t,u_{t-1})$. If we choose as risk factors $%
X_t$ the values of base assets, then $N=M$ and $V_t(X_t,u_{t-1})=
u_{t-1}^{\dagger}X_t$.

\begin{itemize}
 \item \textbf{Ex1:} Total portfolio wealth level at time $t$.
 \item \textbf{Ex2:} Wealth generated by the intraday trading during the period $[t-1,t]$.
\end{itemize}

\item \textbf{Constraints $\mathcal{C}$:}\\ linear equality and inequality constraints
in the rules $u_t$.

\begin{itemize}
 \item \textbf{Ex1:} Self-financing constraint $V_t(X_t,u_{t-1})=V_t(X_t,u_{t})$ for all $t=0,\dots,T-1$ (equality constraint), and lower and upper bounds in the portfolio holdings (inequality constraints).
 \item \textbf{Ex2:} Basin level equations (equality constraints), and lower and upper bounds for water turbined or pumped as well as for basin levels (inequality constraints).
\end{itemize}

\item \textbf{Optimization problem:}\\ given positive deterministic weights $%
(\beta_t)_{t=1,\dots T}>0$ modelling the relative importance assigned to the
measurements in the different subintervals, the optimization problem $(P)$
writes
\begin{equation}  \label{P}
\boxed{
\max_{u\in\mathcal{C}}\mathbb{E}_0\left[\sum_{t=1}^{T}%
\beta_tU(V_t(X_t,u_{t-1}))\right]. }
\end{equation}
\end{itemize}

\begin{rem}
The role of internal states is to simplify the representation of constraints and the recursion formulae, but we could formulate and solve the optimization problem without introducing them. However, they typically are quantities of interest for the problem at hand.
\end{rem}
\begin{rem}
The structure of the objective function in the optimization (\ref{P}) allows for an application of Bellman's equation leading to a decomposition in one step equations with closed or semiclosed solution. This would not work for a generic utility function for the maximization of the expected utility of the cumulated values over the different time subperiods.
\end{rem}

\section{Solution Method}\label{sec3}

\label{Sol} To solve the optimization problem $(P)$ formulated in (\ref{P})
we modify the model of Blomvall and Lindberg described in \cite{BL00}, \cite%
{BL02}, \cite{BL02Bis}, \cite{BL03} and applied in \cite{BL03Bis} by
adapting it to our needs. The constraint set $\mathcal{C}$ can be decomposed
as union of inequality and equality constraints
\begin{equation}
\mathcal{C}=\mathcal{C}_{\text{ineq}}\cup\mathcal{C}_{\text{eq}},
\end{equation}
\noindent where we have set

\begin{itemize}
\item $\mathcal{C}_{\text{ineq}}$: inequality constraints. In our case they
are linear inequalities, which reads
\begin{equation}
\mathcal{C}_{\text{ineq}}:=\{E_tY_t+F_tu_t-e_t\ge0\,|\,t=0,\dots,T-1\}.
\end{equation}
Thereby, $(E_t)_{t=0,\dots,T-1}\subset\mathbf{R}^{L\times M}$,
$(F_t)_{t=0,\dots,T-1}\subset\mathbf{R}^{L\times N}$ and\\ $(e_t)_{t=0,%
\dots,T-1}\subset\mathbf{R}^{L\times 1}$ are processes adapted to the
filtration.

\item $\mathcal{C}_{\text{eq}}$: equality constraints, like the selffinancing condition in Ex $1$ or the basin level equation in Ex $2$ given by the
stochastic dynamics, which reads
\begin{equation}\label{dyn}
\mathcal{C}_{\text{eq}}:=\{Y_{t+1}=g_{t+1}(Y_{t},u_{t},X_{t+1})\,|\,t=0,%
\dots,T-1\}.
\end{equation}
for appropriate choices of the internal states $(Y_t)_t$ and of the functions $(g_t)_t$. The latter typically incorporate the dynamics. Note that $u_{-1}$ denotes the deterministic rule in force just before the rule at
time $0$ is enforced.
\end{itemize}

\text{}\newline
The problem can (but must not) be further simplified by choosing a linear
valuation function and a linear dynamics, that is
$g_t(y,u,x):=A_{t+1}y+B_{t+1}u+b_{t+1}(x)$ and thus
\begin{equation}
\mathcal{C}_{\text{eq}}=\{Y_{t+1}=A_{t+1}Y_t+B_{t+1}u_t+b_{t+1}\,|\,t=0,%
\dots,T-1\},
\end{equation}
where $(A_t)_{t=1,\dots,T}\subset\mathbf{R}^{M\times M}$, $%
(B_t)_{t=1,\dots,T}\subset\mathbf{R}^{M\times N}$ and $(b_t)_{t=1,\dots,T}%
\subset\mathbf{R}^{M\times 1}$ are processes adapted to the filtration.

\text{}\newline
Subsequently, the optimization problem undergoes the following
transformations:

\begin{enumerate}
\item \textbf{($P$):} Original problem (\ref{P}) with a generic concave utility
function $u$ and inequality constraints among others.

\item \textbf{($P_{\mu}$):} Problem with objective function defined as
trade-off between the expected utility and the logarithms of the functions
defining the inequality constraints $u$. The trade-off parameter is denoted
by $\mu>0$. The optimization problem has equality constraints only.

\item \textbf{($\bar{P}_{\mu}$):} Approximation of problem $P_{\mu}$ by
substituting the objective function with its quadratic Taylor polynomial.
\end{enumerate}

\noindent More exactly, we mean that:

\begin{itemize}
\item We write out the expression for the objective function
\begin{equation*}
\mathbb{E}_0\left[\sum_{t=1}^{T}\beta_tU(V_t(X_t,u_{t-1}))\right].
\end{equation*}

\item We extend the objective function by packaging in it all the
restrictions $\mathcal{C}$ utilizing the logarithmic barrier approach, which
approximates the constraints. Thereby, the approximate solution for $(P)$ is
the solution for $(P_\mu)$ for $\mu>0$ small enough.

\item We approximate the extended objective function by its quadratic Taylor
polynomial and the solution of $(P_\mu)$ is given by a Newton's scheme
sequence of solutions of problems of the type $(\bar{P}_{\mu})$.

\item We find optimal rules for the approximated problem (approximated
constraints and approximated objective function).

\item There are two approximations schemes, one for the constraints and one
for the extended objective function. We choose a diagonal sequence to obtain
a sequence of rules converging towards the optimal rules of the original
problem $(P)$.
\end{itemize}

\begin{rem}\label{diff}The differences between this approach and the Blomvall-Lindberg original solution are both formal and substantial:
\begin{itemize}
\item  Blomvall-Lindberg formulate directly the optimization problem on the nodes of a non recombining tree. We formulate it for a general filtration. This is a rather a formal distinction, because the formulae are essentially the same. But it has the advantage of being independent of the way we model the underlying external risk factors. To this aim, conditional expectations are introduced.
\item  The objective function in the Blomvall-Lindberg approach at time $t$ is a function of the risk factors realizations at time $t$. The objective function in our approach at time $t$ is the expectation at time $t$ of the discounted sums of Blomvall-Lindberg's objective functions at times $s=t+1,\dots,T$. In other words, in the case of the hydro optimization of Ex $2$, our model optimizes at every stage $t$ the expected profit till the final horizon $T$ while Blomvall-Lindberg's model optimizes at every stage $t$ the expected profit for the subperiod $[t,t+1]$.
\end{itemize}
\end{rem}
The remainder of this chapter implements the transformation steps described above and culminates in the optimal control rules (\ref{uStar}) for the problem $(\bar{P}_{\mu})$. Readers not interested in the mathematical details can skip directly to subsection \ref{algo sum}.
\subsection{Interior Point Formulation}\label{subsec31}

The problem $(P_{\mu})$ is an approximation of problem $(P)$ by means of the
logarithmic approach, and reads as
\begin{equation}  \label{P_mu1}
\max_{u\in\mathcal{C}_{\text{eq}}}\mathbb{E}_0%
\left[\sum_{t=1}^{T}\beta_tU(V_t(X_t,u_{t-1}))+\mu1^\dagger\sum_{t=0}^{T-1}%
\log(E_tY_t+F_tu_t-e_t)\right],
\end{equation}
\noindent where $1:=[1,\dots,1]^{\dagger}\in\mathbf{R}^{L\times1}$ and $\mu>0
$ is a real parameter.\par As long as we move inside the interior of the feasible set $E_tY_t+F_tu_t-e_t>0$ for all $t=0,\dots,T-1$, the logarithm function is well defined. As soon as we approach to a boundary point, the logarithmic penalty function tends to $-\infty$. This means that, if the maximum is attained, it must be for an interior point, which depends on the parameter $\mu$. For $\mu\rightarrow0^+$ this interior point converges to a point in the feasible set (on the boundary or in the interior), which is the candidate for the solution to the original problem (\ref{P}).\par
If we choose a linear dynamic and a convex utility function, then, by convex
optimization theory (\cite{Mi86},\cite{Ro96} and \cite{Lu97}), the problem
\begin{equation}  \label{P_mu2}
\boxed{
\begin{split}
\max_{\substack{\\Y_{t+1}=A_{t+1}Y_t+B_{t+1}u_t+b_{t+1}\\\\t=0,1,\dots,T-1}}
\mathbb{E}_0&\left[\sum_{t=1}^{T}\beta_tU(V_t(X_t,u_{t-1}))+\right.\\
&\left.\qquad+\mu1^\dagger\sum_{t=0}^{T-1}\log(E_tY_t+F_tu_t-e_t)\right],
\end{split}
}
\end{equation}
\noindent has always a unique solution. As a matter of fact a convex
function over a convex closed domain has always a global minimum. More exactly,
if the sample space $\Omega$ is finite, then existence and uniqueness of the solution directly follows from Kuhn-Tucker's Theorem, see f.i. Theorem 5.6 in \cite{Mi86}.
The general case is proved in Corollary 3.5.1 of \cite{BS96}.

\subsection{Method of Lagrange Multipliers}\label{subsec32}
The problem $(P_\mu)$ in (\ref{P_mu2}) has only linear restrictions, and
can therefore be solved by a closed expression by utilizing the method of Lagrange
multipliers. The Lagrange principal function reads for the Lagrange
multiplier $\lambda=(\lambda_t(\omega))$
\begin{equation}
\begin{split}
\Phi\left(u;\lambda\right):=&\mathbb{E}_0\left[\sum_{t=1}^{T}%
\beta_tU(V_t(X_t,u_{t-1}))+\mu1^\dagger\sum_{t=0}^{T-1}%
\log(E_tY_t+F_tu_t-e_t)+\right. \\
&\quad\left.-\sum_{t=1}^{T}\lambda_t\left(Y_{t+1}-A_{t+1}Y_t-B_{t+1}u_t-b_{t+1})%
\right)\right],
\end{split}%
\end{equation}
and the corresponding Lagrange equations in the unknown optimal process\\
$u=(u_t(\omega))_{t=0,\dots,T-1}$ and unknown optimal Lagrange multiplier $\lambda=(\lambda_t(\omega))$
\begin{equation}  \label{Lag}
\left\{
\begin{array}{ll}
\frac{\partial \Phi}{\partial u_t}\left(u;\lambda\right)=0\quad
(t=0,\dots,T-1) & \\
&  \\
\frac{\partial \Phi}{\partial \lambda}\left(u;\lambda\right)=0. &
\end{array}
\right.
\end{equation}

\subsection{Newton's Scheme}\label{subsec33}
The second equation in (\ref{Lag}) is equivalent to the dynamics (\ref{dyn}) and the first
equation of (\ref{Lag}) can be solved pathwise in $\omega\in\Omega$ for all processes satisfying such dynamics as a restriction.
If we want to find the zeros of the gradient of the objective function by means of Newton's method, then we have to consider its quadratic Taylor polynomial
\begin{equation}
\boxed{
\begin{split}
h_t(y_{\ge t},u_{\ge t}):=\mathbb{E}_t&\left[\sum_{s=t+1}^{T}
\beta_sU(V_s(x_s,u_{s-1}))+\right.\\
&\qquad\left.+\mu1^\dagger\sum_{s=t}^{T-1}
\log(E_ty_t+F_tu_t-e_t)\right],
\end{split}
}
\end{equation}
and to express its gradient with respect to the variables $y_{\ge
t}:=(y_s)_{s\ge t}$ and $u_{\ge t}:=(u_s)_{s\ge t}$ we introduce
\begin{equation}
\boxed{ \begin{split} q^\dagger_t(y_{\ge t},u_{\ge t}):=\nabla_{y_{\ge
t}}h_t(y,u)=\mathbb{E}_t&\left[\sum_{s=t}^{T}\beta_s\nabla_{x_{\ge
t}}U(V_s(x_s,u_{s-1}))+\right.\\
&\left.+\mu
\sum_{t=s}^{T-1}\left(\frac{1}{E_ty_t+F_tu_t-e_t}\right)^\dagger
E_t\right],\\\\ r^\dagger_t(y_{\ge t},u_{\ge t}) :=\nabla_{u_{\ge
t}}h_t(y,u)=&\mathbb{E}_t\left[\sum_{s=t+1}^{T}\beta_s\nabla_{u_{\ge
t}}U(V_s(x_s,u_{s-1}))+\right.\\
&\left.+\mu\sum_{s=t}^{T-1}
\left(\frac{1}{E_ty_t+F_tu_t-e_t}\right)^\dagger F_t\right], \end{split} }
\end{equation}
\noindent where the vector divisions are made componentwise. The Hessian of
the objective function reads
\begin{equation}
\boxed{ \begin{split} Q_t(y_{\ge t},u_{\ge t})&:=\nabla_{y_{\ge
t}}^2h_t(y,u)=\mathbb{E}_t\left[\sum_{s=t+1}^{T}\beta_s\nabla_{x_{\ge
t}}^2U(V_s(x_s,u_{s-1}))+\right.\\ &\qquad\qquad\left.-\mu
\sum_{s=t+1}^{T}E_t^\dagger\diag\left(\frac{1}{E_ty_t+F_tu_t-e_t}\right)^2
E_t\right],\\\\ P_t(y_{\ge t},u_{\ge t})&:=\nabla_{u_{\ge t}}\nabla_{y_{\ge
t}}h_t(y,u)=\\
&\qquad\qquad\qquad=\mathbb{E}_t\left[\sum_{s=t+1}^{T}\beta_s\nabla_{u_{\ge
t}}\nabla_{y_{\ge t}}U(V_s(x_s,u_{s-1}))+\right.\\
&\qquad\qquad\qquad-\left.\mu
\sum_{s=t}^{T-1}E_t^\dagger\diag\left(\frac{1}{E_ty_t+F_tu_t-e_t}\right)^2
F_t\right],\\\\ R_t(y_{\ge t},u_{\ge t})&:=\nabla_{u_{\ge
t}}^2h_t(y,u)=\mathbb{E}_t\left[\sum_{s=t+1}^{T}\beta_s\nabla_{u_{\ge
t}}^2U(V_s(x_s,u_{s-1}))+\right.\\ &\qquad\qquad\left.-\mu
\sum_{s=t}^{T-1}F_t^\dagger\diag\left(\frac{1}{E_ty_t+F_tu_t-e_t}\right)^2
F_t\right]. \end{split} }
\end{equation}
The second order approximation of $h(y,u)$ can be described as a function of
the increment in the variables
\begin{equation}
\begin{split}
\Delta h_t(y_{\ge t},u_{\ge t}):=&h_t(y_{\ge t}+\Delta y_{\ge t},u_{\ge
t}+\Delta u_{\ge t})-h_t(y_{\ge t},u_{\ge t})= \\
&=q_t^\dagger(y_{\ge t},u_{\ge t})\Delta y_{\ge t}+\frac{1}{2}{\Delta
y^{\dagger}_{\ge t}}Q_t(y_{\ge t},u_{\ge t}) \Delta y_{\ge t} + \\
&+r_t^\dagger(y_{\ge t},u_{\ge t})\Delta u_{\ge t}+\frac{1}{2}{\Delta
u^{\dagger}_{\ge t}}R_t(y_{\ge t},u_{\ge t}) \Delta u_{\ge t}+\\
&+{\Delta
y^{\dagger}_{\ge t}}P_t(y_{\ge t},u_{\ge t}) \Delta u_{\ge t},
\end{split}%
\end{equation}
and the matrix
\begin{equation}
\left[
\begin{array}{cc}
Q_t(y,u) & P_t(y,u) \\
P_t^{\dagger}(y,u) & R_t(y,u) \\
\end{array}
\right]
\end{equation}
is positive definite for all $t,y,u$ and $\omega$, an so are the matrices $%
Q_t(y,u)$ and $R_t(y,u)$. The second order expansion of $(P_\mu)$ in (\ref%
{P_mu1}) denoted as $(\bar{P}_\mu)$ is the following quadratic optimization
on $\Omega$
\begin{equation}  \label{Pbar}
\boxed{ \max_{\substack{u=(u_t)_{t=0,\dots,T-1}\\\\\Delta
y_{t+1}=A_{t+1}\Delta y_{t}+B_{t+1}\Delta u_t}}\Delta h_0(y,u), }
\end{equation}
\noindent that is
\begin{equation}  \label{multi}
\begin{split}
\max_{\substack{ u=(u_t)_{t=0,\dots,T-1} \\  \\ \Delta y_{t+1}=A_{t+1}\Delta
y_{t}+B_{t+1}\Delta u_t}} &\left(q_0^\dagger\Delta y+\frac{1}{2}{\Delta y}
^{\dagger}Q_0 \Delta y+r_0^\dagger\Delta u+\frac{1}{2}{\Delta u}
^{\dagger}R_0 \Delta u +\right.\\
&\left.\quad +\frac{1}{2}{\Delta y}^{\dagger}P_0 \Delta u\right).
\end{split}
\end{equation}

\subsection{Dynamic Programming Solution}\label{subsec34}

We solve $(\bar{P}_\mu)$ by dynamic programming and, to this end, we
introduce value functions
\begin{equation}
\begin{split}
J_t(\Delta y_{\ge t}):=&\max_{\substack{ u=(u_s)_{s=t,\dots,T-1}  \\  \\ %
\Delta y_{s+1}=A_{s+1}\Delta x_{s}+B_{s+1}u_s  \\  \\ s=t,\dots,T-1}}\mathbb{%
E}_t\left[q_t^\dagger\Delta y_{\ge t}+\frac{1}{2}{\Delta y_{\ge t}}
^{\dagger}Q_t \Delta y_{\ge t}+\right. \\
&\qquad\qquad\left.+r_t^\dagger\Delta u_{\ge t}+\frac{1}{2}{\Delta u_{\ge t}}^{\dagger}R_t \Delta u_{\ge t} + \frac{1%
}{2}{\Delta y_{\ge t}}^{\dagger}P_t \Delta u_{\ge t}\right],
\end{split}%
\end{equation}
which allow to formulate Bellman's backward recursion as
\begin{equation}
\begin{split}
J_t(\Delta y_{\ge t})&=\max_{\substack{ u_t  \\  \\ \Delta
y_{t+1}=A_{t+1}\Delta y_{t}+B_{t+1}u_t}}\left\{\overline{q}_t^\dagger\Delta
y_{t}+\frac{1}{2}{\Delta y_{t}}^{\dagger}\overline{Q}_t \Delta y_{t}+%
\overline{r}_t^\dagger\Delta u_{t}+\right. \\
&\left.+\frac{1}{2}{\Delta u_{t}}^{\dagger}\overline{R}_t \Delta u_{t} +
\frac{1}{2}{\Delta y_{t}}^{\dagger}\overline{P}_t \Delta u_{t}+\mathbb{E}_t%
\left[J_{t+1}(\Delta y_{\ge t+1})\right]\right\},
\end{split}%
\end{equation}
where
\begin{equation}
\boxed{ \begin{split} \overline{q}_t^\dagger(y_t,u_t)&
:=\nabla_{y_t}h_t(y,u)=\mathbb{E}_t\left[\beta_{t+1}%
\nabla_{y_t}U(V_{t+1}(x_{t+1},u_{t}))+\right.\\
&\left.\quad\qquad\qquad\qquad\qquad+\mu
\left(\frac{1}{E_ty_t+F_tu_t-e_t}\right)^\dagger E_t\right],\\\\
\overline{r}_t^\dagger(y_t,u_t)&
:=\nabla_{u_t}h_t(y,u)=\mathbb{E}_t\left[\beta_{t+1}%
\nabla_{u_t}U(V_{t+1}(x_{t+1},u_{t}))+\right.\\
&\left.\quad\qquad\qquad\qquad\qquad+\mu
\left(\frac{1}{E_ty_t+F_tu_t-e_t}\right)^\dagger F_t\right],\\\\
\overline{Q}_t(y_t,u_t)&:=\nabla_{y_t}^2h_t(y,u)=\mathbb{E}_t\left[%
\beta_{t+1}\nabla_{y_t}^2U(V_{t+1}(x_{t+1},u_{t})))\right.\\
&\left.\quad\qquad\qquad\qquad\qquad-\mu
E_t^\dagger\diag\left(\frac{1}{E_ty_t+F_tu_t-e_t}\right)^2 E_t\right],\\\\
\overline{P}_t(y_t,u_t)
&:=\nabla_{u_t}\nabla_{y_t}h_t(y,u)=\mathbb{E}_t\left[\beta_{t+1}%
\nabla_{u_t}\nabla_{x_t}U(V_{t+1}(x_{t+1},u_{t})))\right.+\\
&\quad\qquad\qquad\qquad\qquad\left.-\mu
E_t^\dagger\diag\left(\frac{1}{E_ty_t+F_tu_t-e_t}\right)^2 F_t\right],\\\\
\overline{R}_t(y_t,u_t)
&:=\nabla_{u_t}^2h_t(y,u)=\mathbb{E}_t\left[\beta_t%
\nabla_{u_t}^2U(V_t(x_t,u_{t})))\right.+\\
&\quad\qquad\qquad\qquad\qquad\left.-\mu
F_s^\dagger\diag\left(\frac{1}{E_ty_t+F_tu_t-e_t}\right)^2 F_t\right],
\end{split} }
\end{equation}
assuming that the matrices $R_t$ and $Q_t$ have the form
\begin{equation}
\begin{aligned} &R_t=\left[ \begin{array}{cc} \overline{R}_t & 0 \\ 0 &
R_{t+1} \\ \end{array} \right] \qquad&Q_t=\left[ \begin{array}{cc}
\overline{Q}_t & 0 \\ 0 & Q_{t+1} \\ \end{array} \right]. \end{aligned}
\end{equation}
This is equivalent with the\newline
\textbf{Inductive Assumption: } $J_t$ is a quadratic function in $\Delta y_t$%
:
\begin{equation}  \label{ind}
J_t(\Delta y_{\ge t})=J_t(\Delta y_{t})=\alpha_t + w_t^\dagger \Delta y_t+%
\frac{1}{2}{\Delta y_t}^\dagger W_t\Delta y_t,
\end{equation}
where $(W_t)_{t=0,\dots,T-1}\subset\mathbf{R}^{M\times M}$ is an adapted,
definite matrix valued process and $(\alpha_t)_{t=0,\dots,T-1}\subset\mathbf{%
R}$, $(w_t)_{t=0,\dots,T-1}\subset\mathbf{R}^{M\times1}$ are adapted
processes.\newline
Using the dynamics $\Delta y_{t+1}=A_{t+1}\Delta y_{t}+B_{t+1}u_t$ we can
rewrite the value function (\ref{ind}) as
\begin{equation}
\begin{split}
J_{t+1}(\Delta y_{\ge t+1})=&\alpha_{t+1} + w_{t+1}^\dagger A_{t+1}\Delta
y_{t}+\frac{1}{2}{\Delta y_{t}}^\dagger A_{t+1}^\dagger W_{t+1}A_{t+1}\Delta
y_{t}+\\
&+\frac{1}{2}{\Delta u_{t}}^\dagger B_{t+1}^\dagger W_{t+1}B_{t+1}\Delta
u_{t}+ \\
&+(w_{t+1}^\dagger+\Delta y_{t}^\dagger A_{t+1}^\dagger W_{t+1})
B_{t+1}\Delta u_t.
\end{split}%
\end{equation}
With the definitions
\begin{equation}  \label{R1}
\boxed{ \begin{aligned}
\widetilde{a}_t&:=\sum_{s=t+1}^{T}\alpha_s\qquad&\widetilde{r}_t:=r_t+%
\sum_{s=t+1}^{T}B_s^{\dagger}w_s^{\dagger},\\
\widetilde{R}_t&:=R_t+\sum_{s=t+1}^{T}B_s^{\dagger}W_sB_s\qquad&%
\widetilde{q}_t:=\overline{q}_t+\sum_{s=t+1}^{T}A_s^\dagger w_s,\\
\widetilde{Q}_t&:=\overline{Q}_t+\sum_{s=t+1}^{T}A_s^\dagger W_s
A_s\qquad&\widetilde{P}_t:=\overline{P}_t+\sum_{s=t+1}^{T}A_s^\dagger W_s
B_s, \end{aligned} }
\end{equation}
expression(\ref{ind}) for the value function becomes
\begin{equation}  \label{ind2}
\begin{split}
J_t(\Delta y_{t})=\max_{\Delta u_t}&\left[\widetilde{a}_{t} + \widetilde{q}%
_t^{\dagger}\Delta y_{t}+\frac{1}{2}{\Delta y_{t}}^\dagger \widetilde{Q}%
_t\Delta y_{t}+\left(\widetilde{r}_t^\dagger+ {\Delta y_{t}}^\dagger
\widetilde{P}_t\right)\Delta u_{t} +\right.\\
&\quad+\left.\frac{1}{2}{\Delta u_{t}}^\dagger
\widetilde{R}_t\Delta u_{t}\right].
\end{split}
\end{equation}

\subsection{Optimal Control Rules}\label{subsec35}

The optimum can be found by differentiating the expression maximized in (\ref%
{ind2}) with respect to $\Delta u_t$:
\begin{equation}
\begin{split}
0&=\nabla_{\Delta u_t}\left[\widetilde{a}_{t} + \widetilde{q}%
_t^{\dagger}\Delta y_{t}+\frac{1}{2}{\Delta y_{t}}^\dagger \widetilde{Q}%
_t\Delta y_{t}+\left(\widetilde{r}_t^\dagger+ {\Delta y_{t}}^\dagger
\widetilde{P}_t\right)\Delta u_{t} +\right.\\
&\qquad\qquad+\left.\frac{1}{2}{\Delta u_{t}}^\dagger
\widetilde{R}_t\Delta u_{t}\right]=\widetilde{r}_t^\dagger+{\Delta y_{t}}^\dagger \widetilde{P}_t+ {\Delta
u_{t}}^\dagger \widetilde{R}_t,
\end{split}%
\end{equation}
which means, being $R_t$ symmetric,
\begin{equation}  \label{uStar}
\boxed{\Delta
u_t^*=-\widetilde{R}_t^{-1}(\widetilde{r}_t+\widetilde{P}_t^\dagger \Delta
y_{t}).}
\end{equation}
Inserting this optimal $\Delta u_t^*$ in (\ref{ind2}), the value function
becomes
\begin{equation}
J_t(\Delta x_{t})=\alpha_{t} + w_t^{\dagger}\Delta x_{t}+\frac{1}{2}{\Delta
x_{t}}^\dagger W_t\Delta x_{t},
\end{equation}
where
\begin{equation}  \label{R2}
\boxed{ \begin{split}
\alpha_t&:=\widetilde{a}_t-\frac{1}{2}\widetilde{r}_t^\dagger
\widetilde{R}_t^{-1}\widetilde{r}_t,\\
w_t&:=\widetilde{q}_t-\widetilde{P}_t\widetilde{R}_t^{-1}\widetilde{r}_t,\\
W_t&:=\widetilde{Q}_t-\widetilde{P}_t\widetilde{R}_t^{-1}\widetilde{P}_t^%
\dagger, \end{split} }
\end{equation}
The expression for $W_t$ in the third equation of (\ref{R2}) together with (%
\ref{R1}) is known as the \textbf{discrete time Riccati equation} in control
theory.
\begin{rem}
If $W_{t}$ is positive definite, if $W_s$ is positive semidefinite for all $s=t+1,\dots,T$.
\end{rem}

\subsection{The Algorithm}\label{algo sum}
Newton's step determination problem $(\bar{P}_\mu)$ in (\ref{Pbar}) for the
barrier subproblem $(P_\mu)$ is solved by (\ref{uStar}), where matrices,
vectors and constants are defined recursively by (\ref{R1}) and (\ref{R2}).
\begin{figure}[!]
\centering
\includegraphics[width=15cm, angle = -90]{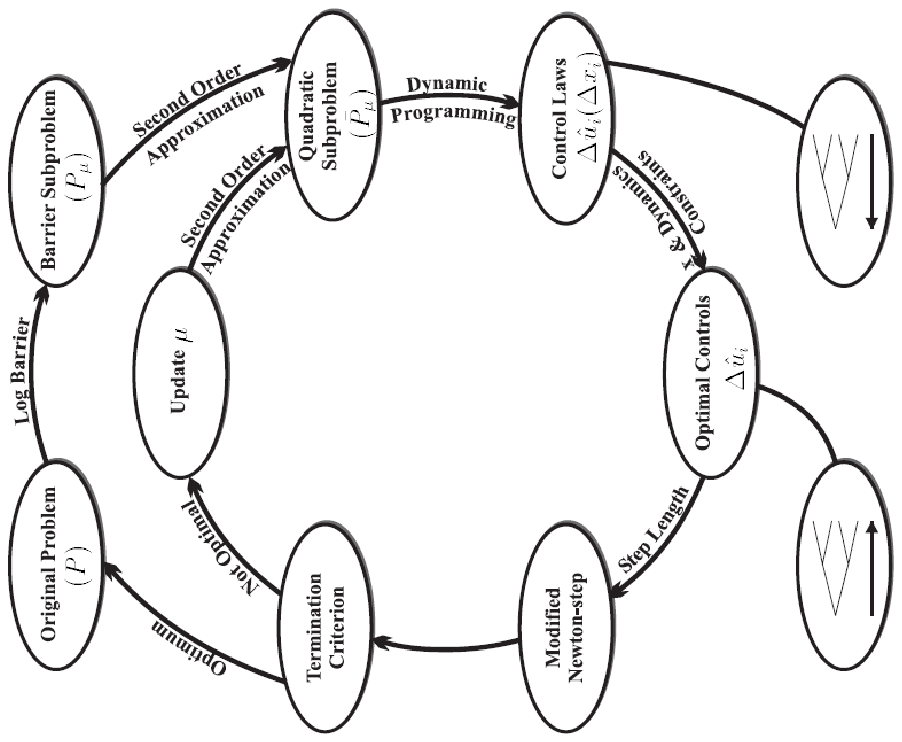}\newline
\caption{Converging Sequence of Optimal Rules (Picture from \protect\cite%
{BL02})}
\label{diag}
\end{figure}

\section{Implementation}\label{sec4}
The purpose of this section is to show that the solution algorithm shown in Figure \ref{diag} can be efficiently implemented by means of time and space discretization on a lattice, i.e. a particular kind of tree, where each node will host a realization of risk drivers, risk factors and optimal rules, comparing this modelling choice with other possible approaches. When we want to implement discrete time dynamic stochastic programming models, we have basically four possibilities:

\begin{enumerate}
\item \textbf{The (semi-)closed formula solution:}\newline
In some (seldom) cases one can find a set of (semi-)closed formulae
representing the optimal control rules as a functional of conditional
expectations of functions of risk factors. The optimal rules can be
therefore explicitly determined given a probability model for the risk
factors. But, in most of the cases, the computation can be only numerical,
and we therefore have to switch to

\item \textbf{The graph solution:}\newline
There are several possibilities to choose a graph, and for all the nodes of
the graph will have to correspond to the atoms of the sigma algebras of the
filtration $(\mathcal{A}_t)_{t=0,\dots T}$:

\begin{enumerate}
\item \textbf{The full non recombining tree:}\newline
This is the most generic solution, which has the disadvantage of being
implementable in its fully fledged version on high performing computer only,
because the number of nodes in a time layer increases exponentially with
time. The alternative is to reduce drastically the number of branches from
every node when time increases. To do so, one has to develop criteria to
generate representative branches. Those criteria are mostly heuristic.

\item \textbf{The grid:}\newline
Parallel paths for simulated external states are stored in the nodes. If we
have a (semi-) closed formula for the optimal rules, these can be computed
on every node. If not, then the optimal rule is computed on the node by
solving the Bellman's backward optimization step by simulating jumps from
that node to all nodes in the following time layer. This method is
computationally effective and is therefore widespread.

\item \textbf{The lattice:}\newline
We see a lattice with many branches as a totally recombining tree.
Therefore, being the number of nodes in a time layer a linear function of
time, the full fledged model is implementable even on standard computers. Of
course, the main challenge is to fill the nodes with state realizations in
such a way that these are compatible with their dynamics on one hand, and
with the full recombining property of the graph, on the other. To our
knowledge this method is new, and is a generalization of binomial and
trinomial trees' construction utilized for option pricing. This is the way
we choose here. It has the advantage of being extensible to the case where
no (semi-)closed solution on the nodes exists, and is thus a viable
implementation method for a numerical solution of Bellman's backward
recursion. In contrast to the grid method one does not have to resimulate the jumps from one node into its children everytime
the algorithm performs an approximation step.
\end{enumerate}
\end{enumerate}
The algorithm for the lattice construction, the simulation of states and the approximation of optimal control rules is structured
into the following steps:

\begin{description}
\item[\textbf{Step 1}] \text{}\\ We construct the lattice with $k$ branches for every node and final
horizon $T$. Let $e_s:=(0,\dots,1,0,\dots,0)\in\mathbf{R}^k$ is the $s$th standard basis vector. For $t=0,\dots,T$ we set
\begin{equation}
\boxed{
\begin{split}
&\mathcal{L}:=\bigcup_{t=0}^T\mathcal{L}_t:\text{lattice},\\
&\\
&\mathcal{L}_t:=\left\{(t,i)\in\{0\dots,T\}\times\mathbf{N}_0^k\left|\sum _{s=1}^ki_s=t\right.\right\}:\text{lattice time $t$ layer},\\
&\\
&n_t(i)=(t,i)\in\mathcal{L}_t:\text{node layer at time $t$}\\
&\\
&\text{Children}(n_t(i)):=\{(t+1,i+e_s)\left|\,s=1,\dots,k\right.\},\\
&\\
&\text{Parents}(n_s(j)):= \{(s-1,i)\left|\,n_s(j)\in\text{Children}((s-1,i))\right.\},\\
&\\
&\text{Parents}(n_0(0)):=\{\},
\end{split}
}
\end{equation}
\noindent where the number of nodes at time $t$ is
\begin{equation}
\boxed{N_t:=\left|\mathcal{L}_t\right|=(k-1)t+1=O(t),}
\end{equation}
\noindent and the total number of nodes is
\begin{equation}
\boxed{\mathcal{N}_T:=\left|\mathcal{L}\right|=\sum_{t=0}^TN_t=\left((k-1)%
\frac{T}{2}+1\right)(T+1)=O(T^2).}
\end{equation}
Geometrically speaking the (infinite) lattice consists in the points in the $k$-dimensional space with non negative integer coordinates.
The time $t$ layer of the lattice consists in the points laying on the hyperplane with orthogonal vector $(1,1,\dots,1)$ passing through the point $(k,0,\dots,0)$.\\
\item[\textbf{Step 2}] \text{}\\ We introduce the probability space $(\Omega, P,\mathcal{A})$, where
the cartesian product
\begin{equation}
\boxed{ \Omega:=\text{X}_{t=0}^T\,\text{Children}^t(n_0(0)), }
\end{equation}
corresponds to all possibilities of traveling across the lattice from left
to right as times goes by,
\begin{equation}
\boxed{ \mathcal{A}:=\mathcal{P}(\Omega) }
\end{equation}
is the sigma algebra of all measurable events, and, the sigma algebra
generated by the lattice nodes in the time layer $t$, (that is having the
nodes as basis) leads to a filtration $\left(\mathcal{A}_t\right)_{t=0,%
\dots,T}$, where
\begin{equation}
\boxed{ \mathcal{A}_t:=\sigma\left(\mathcal{L}_t\right). }
\end{equation}
The probability of every node event is recursively defined as:
\begin{equation}
\boxed{ \begin{split} P[n_t]&:=
\sum_{n_{t-1}\in\,\text{Parents}(n_t)}\frac{P[n_{t-1}]}{k}\\ P[n_0]&:=1,
\end{split} }
\end{equation}
and that for the elementary event $\omega=\left(n_0(0),n_1(i_1),%
\dots,n_{T-1}(i_{T-1}),n_{T}(i_{T})\right)$ is
\begin{equation}
\boxed{ P[\omega]:= \frac{1}{k^T}, }
\end{equation}
\item[\textbf{Step 3}] \text{}\\ By means of simulations we fill the lattice nodes with realizations of
the risk drivers $Z=(Z_t)_{t=1,\dots,T}$. Since these are multivariate
i.i.d. over time, these simulations are straightforward: for every $%
t=1,\dots,T$

\begin{enumerate}
\item simulate $N_t$ values $z_t^1,\dots,z_t^{N_t}$ values with all the same
probability.

\item set $Z_t(n_t^k):=z^k_t$ for all nodes in $\mathcal{L}_t$, the layer at
time $t$.
\end{enumerate}

These are simulated values for the risk drivers on the nodes.\\

\item[\textbf{Step 4}] \text{}\\  We compute the corresponding realizations of the external states (risk
factors) $X=(X_t)_{t=1,\dots,T}$, by translating the dynamics at elementary
event level $X_t(\omega):=f_t(X_{t-1}(\omega),Z_t(\omega))$ to the nodes as
\begin{equation}
X_{t+1}(\overline{n}):=\sum_{n\in\text{Parents}(\bar{n})}\frac{p(n)}{%
\sum_{n\in\text{parents}(\bar{n})}p(n)}f_{t+1}(X_{t}(n), Z_{t+1}(n)).
\end{equation}

\item[\textbf{Step 5}] \text{}\\  We pick a $\mu=\mu_{\text{Start}}>0$ and a positive sequence $%
(\mu_j)_{j\ge0}$ such that $\mu_0=\mu_{\text{Start}}$ and $%
\mu_j\rightarrow0^+$ $(j\rightarrow+\infty)$.\\

\item[\textbf{Step 6}] \text{}\\  We pick initial values for the control variables $u_t$ and the
internal states $y_t$.\\

\item[\textbf{Step 7}] \text{}\\  For the value $\mu$ we compute all the realization of the processes in
(\ref{R1}) and (\ref{R2}) by inserting the realizations of all control rules
and both internal and external states.\\

\item[\textbf{Step 8}] \text{}\\  We compute $\Delta u_t^*$ and check if it is approximatively very
small. If not then, for $\Delta y_t(\Delta u_t^*)$ and do the increase step
\begin{equation}
\begin{split}
&u_t\mapsto u_t + \Delta u_t^* \\
&y_t\mapsto y_t + \Delta y_t(\Delta u_t^*),
\end{split}%
\end{equation}
update $\mu$ according to the sequence in (5) and jump to point $(7)$. If $%
\Delta u_t^*$ is too big, so that $u_t + \Delta u_t^*$ and $y_t + \Delta
y_t(\Delta u_t^*)$ lie outside the feasible set, then $\Delta u_t^*$ has to
be substituted by $\epsilon_t\Delta u_t^*$ for an appropriate $%
\epsilon_t\in]0,1[$ small enough. Typically, $\epsilon_t$ depends on the
node where it is computed.
\end{description}

\begin{rem}There are different possibilities to choose $\epsilon_t$ to guarantee feasibility. Blomvall and Lindberg propose to choose the same $\epsilon_t$ for all $t$ and all nodes by looking at the largest $\overline{\epsilon}$ such that $u_t+\epsilon\Delta u_t^*$ is still feasible for all nodes and all times, and then set $\epsilon:=\min(\xi\overline{\epsilon},1)$ for a $\xi\in]0,1[$. We, instead, proceed layerwise. Assuming that up to time layer $t-1$ the appropriate choice has already being made, in order to find a node dependent $\epsilon_t$ for all nodes in the time layer $t$ we look for a node dependent $\overline{\epsilon}_t$ such that $u_t+\overline{\epsilon}_t\Delta u_t^*$ and $u_s$ are feasible for all $s=t+1,\dots,T-1$. This can be efficiently achieved by a linear program, where the objective function is not really relevant. For a fixed $\xi\in]0,1[$ we then set $\epsilon_t:=\min(\xi\overline{\epsilon}_t,1)$ for all nodes in the layer, and repeat the procedure for the next time step. This refined procedure guarantees a faster convergence then Blomvall and Lindberg's when the maximizer lies on the boundary of the feasible set.
\end{rem}
\begin{rem} Why does the implementation on the lattice work? When implementing the dynamics, there is a fundamental difference between the non recombining tree and the lattice. The value of a process on a node depends on the values of the process on the parent nodes. In the non recombining tree case a node has only one parent, while in the lattice case a node can have several parents. But in both cases the process values on the nodes are expressed by conditional expectations. More exactly, we have the situation summarized in Table \ref{LV}.
\begin{table}[!]
\begin{tabular}{|l|l|l|}
\hline
\textbf{Symbol}&\textbf{Description} & \textbf{Mapped to} \\ \hline\hline
$\Omega$ & Space of all  & All possibilities of travelling \\
& elementary events& through the lattice from left to right\\ \hline
$n_t$ & Node & An atom $\mathcal{B}(n_t)$ of the $\sigma$-algebra \\
& &for the time layer $t$ containing that node\\ \hline
$Y(n)$ &Value on the node $n$  & $Y(n)=\mathbb{E}[Y|\mathcal{B}(n)]\neq Y(\omega)\quad\text{ for }\omega\in n$ \\
 &of any random variable $Y$ &\\ \hline
$X_t(n)$ &Value on the node $n$  &$X_t(n)=\mathbb{E}[X_t|\mathcal{B}(n)]$\\
& of the external state  $X_t$ &\\ \hline
\end{tabular}
\caption{Lattice Variables}
\label{LV}
\end{table}

\noindent An internal state variable defined as $Y_{t+1}=g_{t+1}(Y_{t},u_{t},X_{t+1})$ for deterministic functions $g_t$ for $t=1,\dots,T$, typically utilized to define constraints. On the nodes it is represented by   $Y_t(n)=\mathbb{E}[Y_t|\mathcal{B}(n)]$, and at elementary event level it has the dynamics
    \begin{equation}
    Y_{t+1}(\omega)=g_{t+1}(Y_{t}(\omega),u_t(\omega),X_{t+1}(\omega)),
    \end{equation}
    which becomes the external state variable dynamics at node level
    \begin{equation}
    \begin{split}
     Y_{t+1}(\overline{n})&:=\mathbb{E}[Y_{t+1}|\mathcal{B}(\bar{n})]=\sum_{n\in\text{Parents}(\bar{n})}\frac{p(n)}{p(\bar{n})}\mathbb{E}[Y_{t+1}|\mathcal{B}(n)]=\\
     &=\sum_{n\in\text{Parents}(\bar{n})}\frac{p(n)}{\sum_{n\in\text{parents}(\bar{n})}p(n)}g_{t+1}(Y_{t}(n),u_t(n),X_{t+1}(n)).
    \end{split}
    \end{equation}
   This holds for a generic dynamics of the internal states and hence for the implemented linear dynamics $g_t(y,u,x):=A_{t}y+B_tu+b_t(x)$.

\end{rem}

\section{Application: Water Values and Intraday Electricity Trading}\label{sec5}

\label{sec5} The algorithm presented in the preceding section can be
utilized to optimize intraday electricity trading and model at the same time
water values for hydro assets.

Everyday by 11:00 CEST all the participants to the Swiss electricity spot
market have to submit to the energy exchange their aggregated bids for the
day-ahead both demand and supply. These, in the ''ask''-case specify for
every hour of the following day, from 00:00 till 24:00$^-$ CEST the quantity
of energy $\Xi_t^\text{Ask}$ in MWh that one participant is willing to
deliver during that hour $t=0,\dots,23$ if the electricity price $S_t$ then
is greater than or equal to a certain value $\text{GP}_t^\text{Ask}$, called
\textit{generation water value}. In the ''bid''-case the electric market
participants specify for every hour of the following day the quantity of
energy $\Xi_t^\text{Bid}$ in MWh that the participant is willing to buy
during that hour $t=0,\dots,23$ if the electricity price $S_t$ then is
smaller than or equal to a certain value $\text{GP}_t^\text{Bid}$, called
\textit{delivery water value}. For every hour the energy exchange aggregates
all asks and all bids two monotone step functions, the ask curve and the bid
curve, representing the quantity of energy deliverable (ask) or requested
(bid) as a function of the price. The intersection point of the two curves,
i.e. the market clearing price at time $t$ is the spot price which will hold
for the hour $t$ of the next day. The $24$ spot prices for the day-ahead are
published at around 11:15 CEST of the current day. Note that all of the
market participants are due to deliver or to buy the quantities of energy
specified during the bidding process, but at the market clearing price
determined by the energy exchange for the day-ahead spot prices. However,
the auction is not physically binding, that is, energy must not necessarily
be produced but can be bought and delivered.

All the trades for the day ahead settled between 11:15 and 23:59 CEST, where
energy quantities $\Xi_t^{\text{Spot, Sell}}$ and $\Xi_t^{\text{Spot, Buy}}$
will be sold and respectively bought at hour $t$ of the next day at price $%
S_t$ have to be taken into account by the trading strategy of the intraday -
given what the spot desk has done. Given a certain utility function $U:\mathbf{R}\hookrightarrow \mathbf{R}$, the relevant optimization problem at
23:59 CEST of the day before $(t=0$) reads for $T:=24$ and $t_0:=1$
\begin{equation}  \label{Opt1}
\max_{\substack{ (u_t)_{t=t_0,\dots,T-1} \\ \text{Restrictions}}}\mathbb{E}_0%
\left[\sum_{t=t_0}^{T}\beta_tU(V_t(X_t,u_{t-1}))\right]
\end{equation}
and we make the choices needed to model the intraday-spot $P\&L$ in
\begin{itemize}
\item $\beta_t:=1$ for all $t$,

\item $X_t$ is the intraday price holding during $]t-1,t]$,

\item We assume that we have $B$ basins, labelled with $b=1,\dots,B$. Basin $%
b$ is connected with $N_b$ turbines/pumps. Turbine/pump $j_b=1,\dots,N_b$
processes $u_t^{b,j_b}$ energy at time $t$. The aggregated processed energy
quantity at time $t$ for basin $b$ is given by $u_t^b:=%
\sum_{j_b=1}^{N_b}u_t^{b,j_b}$ and for the whole hydro infrastructure by $%
u_t:=\sum_{b=1}^Bu^b_t$.

\item $V_t(X_t,u_{t-1}):=u_{t-1}X_t+(\Xi_t^{\text{Spot, Sell}}-\Xi_t^{\text{%
Spot, Buy}})S_t$ is the portfolio profit and loss for both spot and intraday
desks.
\end{itemize}

\noindent The optimization problem reads after these choices
\begin{equation}  \label{Opt2}
\max_{\substack{ {(u_t)_{t=t_0,\dots,T-1}} \\ {\text{Restrictions}}}} \;%
\mathbb{E}_0\left[\sum_{t=t_0}^{T}U(u_{t-1}X_t+(\Xi_t^{\text{Spot, Sell}}-\Xi_t^{\text{%
Spot, Buy}})S_t)\right].
\end{equation}
\noindent The restrictions are listed in Table \ref{Restr} and explained in detail here below.
\begin{table}[!]
\begin{tabular}{|l|l|}
\hline
\textbf{Restriction}&\textbf{Description}  \\ \hline\hline
Hydro-infrastructure dynamics   & Equations connecting basin levels \\
 & and water inflows or outflows\\ \hline
Lower and upper bounds  & Limits for turbines and pumps\\
for the energy produced &  \\ \hline
\end{tabular}
\caption{Restrictions}
\label{Restr}
\end{table}

\begin{itemize}
\item \textbf{The dynamics of the hydro infrastructure} connecting:

\begin{itemize}
\item the basins' volumes,

\item the water inflows,

\item the water outflows (turbined water, overspills).
\end{itemize}

The basin $b$ level dynamics $(Y_t^b)_{t=t_0,\dots,T-1}$ is given for all $%
t=t_0,\dots,T-1$ by
\begin{equation}  \label{level}
Y^b_{t+1}(\omega)=Y^b_{t}(\omega)-u^b_t(\omega)+i^b_{t+1}(\omega)
\end{equation}
where the process $(i^b_t)_{t=t_0+1,\dots,T}$ denotes the exogenous dynamics
of basin $b$ inflow, and the level lower and upper constraints are given for
all $t=t_0,\dots,T-1$ by
\begin{equation}  \label{basinconstr}
Y^{b,\text{Min}}\le \mathbb{E}_{t}[Y^b_{t+1}]\le Y^{b,\text{Max}},
\end{equation}
for specified constants $Y^{b,\text{Max}}>Y^{b,\text{Min}}>0$ which are
(flexible) basin characteristics. Remark, that, being the basins' inflows
uncertain, we cannot express (\ref{basinconstr}) as a predictable constraint
for the water turbined or pumped, but the consequences on the optimal
solution are typically not material, because the inflow volatility is small
and we can assume for most applications that the inflow is deterministic and
given as a table characterizing the basins' system.

In contrasts to financial applications we do not have here the self
financing constraint, because we can decide to turbine/pump or not in a
certain period independently of what has been done before or what will be
done afterwards, as long as the basin constraints are not binding.

\item \textbf{Lower and upper bounds for the energy produced by each turbine every
hour}. Note that negative lower bounds account for \textit{pumping}. These
bounds capture expected potential market liquidity restrictions in the day
ahead market and, for all $t=t_0,\dots,T-1$, $j_b=1,\dots,N_b$, $b=1,\dots,B$%
, read as:
\begin{equation}
u^{b,j_b,\text{Min}}\le u^{b,j_b}_t(\omega)\le u^{b,j_b,\text{Max}}.
\end{equation}
\end{itemize}
Finally we make the following modeling choices for the intraday price stochastic dynamics:
\begin{equation}\label{GBM}
dX_t=X_t[\mu_t(X_t)dt+\sigma_t(X_t)dW_t],
\end{equation}
where $\mu_t:\mathbf{R}\rightarrow\mathbf{R}$ and $\sigma_t:\mathbf{R}%
\rightarrow\mathbf{R}^{K\times q}$ are functions with appropriate regularity and $%
(W_t)_{t\ge0}$ is a $K$-dimensional standard Brownian motion with respect to
the statistical measure $P$. We assume that, for the short future period,
the intraday price dynamics is approximatively driftless, i.e. $\mu_t\doteq0$.
 We can assume one risk driver (i.e. $K:=1$) and a deterministic
volatility, that is
\begin{equation}
\sigma_t(X_t(\omega))\equiv\sigma_t\in\mathbf{R}.
\end{equation}
A better way to model intraday prices $X_t$ is by modelling their spreads $%
Z_t:=X_t-S_t$ to spot prices $S_t$
\begin{equation}\label{STS}
dZ_t=Z_t[\nu_t(Z_t)dt+\eta_t(Z_t)dW_t],
\end{equation}
where $\nu_t:\mathbf{R}\rightarrow\mathbf{R}$ and $\eta_t:\mathbf{R}%
\rightarrow\mathbf{R}$ with appropriate regularity. Again, the spread
dynamics is approximatively driftless, i.e. $\nu_t\doteq0$ and we assume a
deterministic volatility, that is
\begin{equation}
\eta_t(Z_t(\omega))\equiv\eta_t=\sqrt{\sigma_t^2+\chi_t^2+2\rho_t\sigma_t%
\chi_t}\in\mathbf{R},
\end{equation}
where $\chi_t$ denotes the instantaneous volatility for the log return of
spot prices, and $\rho_t$ the correlation between log return of spot and
intraday prices. Note that to model intraday prices via their spread to spot
one needs a spot price model first. In particular one has to model the
expected spot prices in the day ahead market.\par

Now we proceed to model water values for the hydro infrastructure described so far.
Before 11:00 CEST we can utilize (\ref{Opt2}) to determine the
generation water values $\text{GP}_t^\text{Ask}$ for $t=0,\dots,23$ for the
day ahead for the hydro infrastructure, whose bids we will aggregate in our
bid for the energy exchange. We exclude for the moment the spot desk whose
trades for the day ahead have not been established yet from (\ref{Opt2}). We
define the water values as the shadow prices associated to the basin levels
dynamics (\ref{level}), that is the value of the Lagrangian multipliers
associated to (\ref{level}) for the optimal solution: they represents the
instantaneous change per unit of constraints (\ref{level}), in [MWh], in the
objective function value of (\ref{Opt2}), in [EUR], for a variation of the
constraints, i.e. the marginal utility of relaxing the basin level
constraints.
\noindent Therefore, after having expressed the basin level dynamics (\ref%
{level}) with the equivalent expression
\begin{equation}  \label{weightedlevel}
p(\omega)[Y^b_{t}(\omega)-Y^b_{t-1}(\omega)+u^b_{t-1}(\omega)-i^b_t(%
\omega)]=0,
\end{equation}
\noindent for all $t=1,\dots,T$ and $b=1,\dots B$, we obtain a Lagrangian
principal function for the basin constraints
\begin{equation}
\begin{split}
\Phi(u,\lambda):=\sum_{\substack{ \omega\in\Omega \\ t=1,\dots,T \\ %
b=1,\dots,B}}p(\omega)&\left[U(u_{t-1}(\omega)X_t(\omega))-\lambda_{t}^b(%
\omega)(Y^b_{t}(\omega)-Y^b_{t-1}(\omega)+\right.\\
&\left.\quad+u^b_{t-1}(\omega)-i^b_{t}(\omega))\right],
\end{split}
\end{equation}
where $u=(u_t^{b,j_b}(\omega))$ is the energy corresponding to the water
turbined or pumped and $\lambda=(\lambda_t^b(\omega))$ is the set of
Lagrangian multipliers for the basin levels. The optimal solution satisfies
the equations
\begin{equation}
\left\{
\begin{array}{ll}
\frac{\partial \Phi(u,\lambda)}{\partial u_{t-1}^b(\omega)}%
=p(\omega)[X_t(\omega)U^{\prime}(u_{t-1}(\omega)X_t(\omega))-\lambda_t^b(%
\omega)]=0 &  \\
\frac{\partial \Phi(u,\lambda)}{\partial \lambda_t^b(\omega)}%
=-(Y^b_{t+1}(\omega)-Y^b_{t}(\omega)+u^b_t(\omega)-i^b_t(\omega))=0, &
\end{array}
\right.
\end{equation}
which leads to
\begin{equation}
\lambda_t^b(\omega)=X_t(\omega)U^{\prime}(u_{t-1}(\omega)X_t(\omega)).
\end{equation}
The choice of the reformulation (\ref{weightedlevel}) takes the probability
for the constraint to be binding into account and leads to the meaningful
definition for the shadow price. Therefore, the stochastic water values are
the same for all basins in the hydro infrastructure and read
\begin{equation}  \label{stochgp1}
\boxed{\text{gp}_t^{\text{Ask}}(\omega):=X_t(\omega)U^{\prime}(u_{t-1}^*(%
\omega)X_t(\omega)) , }
\end{equation}
where $u^*$ is the solution of the optimization problem (\ref{Opt2})
satisfying \textit{all constraints, both equality and inequality ones}. We
can use these stochastic water values to define production water values for
the bid, by taking as possible definition the certainty equivalent of $\text{gp}_t^\text{Ask}$:
\begin{equation}
\boxed{
\text{GP}_t^\text{Ask}:=U^{-1}\mathbb{E}_0\left[U(\text{gp}_t^\text{Ask})\right]. }
\end{equation}
Being the utility function $U$ monotone increasing and concave the risk add on  $U^{-1}\mathbb{E}_0\left[U(\text{gp}_t^\text{Ask})\right]-\mathbb{E}_0\left[\text{gp}_t^\text{Ask}\right]$ is non negative and accounts for the risk aversion.\par
If the initial basin levels are distant enough from the lower and upper
bounds, then we can assume that during the 24 hours of the optimization
interval the basin level constraints are not binding and thus
\begin{equation}
u_t^{b,j_b,*}(\omega)\equiv u^{b,j_b,\text{Max}}.
\end{equation}
To our knowledge the expression ``water value'' was introduced for the first
time by Larsson and Stage in \cite{LS61}. For a treatment of water values
defined by means of Lagrangian multipliers in a cost minimization problem
see \cite{DR04} and an approach consisting in a time dependent shadow
pricing of water in profit maximization problem can be found in \cite{HW07}.

We can consider the joint intraday and spot desks in the determination of
water values. The joint optimization problem at a certain hour before 11:00
CEST (t=0) reads for $T:=48$ and $t_0:=24$
\begin{equation}  \label{Opt3}
\max_{\substack{ (u_t)_{t=t_0,\dots,T-1} \\ (\Xi_t)_{t=t_0,\dots,T-1} \\ {%
\text{Restrictions}}}} \;\mathbb{E}_0\left[\sum_{t=t_0}^{T}U(u_{t-1}X_t+%
\Xi_{t-1}S_t)\right],
\end{equation}
where $(S_t)_{t=t_0+1,\dots,T}$ denote the (till 11:15 CEST) stochastic spot
prices for the day ahead, and $(\Xi_{t})_{t=t_0,T-1}$ the stochastic
quantities of energy turbined for the spot market. The restrictions are
those of (\ref{Opt2}), where $u_t$ is substituted by $u_t+\Xi_{t}$. A
computation analogous to the one for (\ref{stochgp1}) leads to the following
stochastic and deterministic water values for all basins in the hydro power
plant:
\begin{equation}
\boxed{ \begin{split}
\text{gp}_t^{\text{Ask}}(\omega)&:=\frac{1}{2}(X_t(\omega)+S_t(\omega))U^{%
\prime}(u_{t-1}^*X_t(\omega)+\Xi_{t-1}^*(\omega)S_t(\omega))\\
\text{GP}_t^\text{Ask}&:=U^{-1}\mathbb{E}_0\left[U(\text{gp}_t^\text{Ask})\right],
\end{split} }
\end{equation}
where $u^*, \Xi^*$ is the solution of the optimization problem (\ref{Opt3})
satisfying \textit{all contraints, both equality and inequality ones}. As in
the intraday case, if the initial basin levels are distant enough from the
lower and upper bounds, then we can assume that during the 24 hours of the
optimization interval the basin level constraints are not binding and thus
\begin{equation}
u_t^{b,j_b,*}(\omega)+\Xi_t^{b,j_b,*}(\omega)\equiv u^{b,j_b,\text{Max}},
\end{equation}
for all basins and turbines.
\begin{rem}[\textbf{Strategy Extension: Accounting for Hourly Forward Trades}]
The models (\ref{Opt2}) and (\ref{Opt3}) can be utilized at any hour $0,\dots,11$ of the current day to find stochastic and deterministic water values for the hours $\{1,\dots,24\}+24$ of the day ahead. Immediately after 11:15 CEST the day ahead spot prices are known. At any hour $\{11,\dots 48\}$ it is possible to initiate forward transactions with one hour in $\{0,\dots,23\}+24$ as delivery period. This means, at time $t$ of the day ahead the (deterministic) energy quantity $\Psi_t$ will be delivered for the price $F_t$ established when the transaction was closed. In  order for the allocation strategy to take this aspect into account, we choose $T:=48$  and $t_0:=24$ and modify the optimization model (\ref{Opt2}) to
\begin{equation}\label{Opt4}
\max_{\substack{(u_t)_{t=t_0,\dots,T-1}\\{\text{Restrictions}}}}
\;\mathbb{E}_0\left[\sum_{t=t_0}^{T}U(u_{t-1}X_t+\Psi_{t-1}F_t)\right],
\end{equation}
where $(F_t)_{t=t_0,\dots,T}$ denote the deterministic forward prices for the day ahead, and $(\Psi_{t})_{t=0,T-1}$ the deterministic quantities of energy turbined for the forward market, established at a certain hour ($t=0$) of the day before. Of course one can add $(\Psi_t)_{t=t_0,\dots,T-1}$ to the optimization variables and run the at time $\{12,\dots,23\}$ the algorithm solving (\ref{Opt4}) is to find both optimal rules for the turbined quantities in the intraday market in the day ahead and deterministic optimal forwards for the day ahead. From the equality
\begin{equation}
\Psi^\text{*}_t=\Psi^\text{In Force}_t+\Delta \Psi^\text{Forward,*}_t,
\end{equation}
one reads off the energy quantity $\Delta \Psi^\text{*}_t$ to be hedged with the new forward transaction at time $t_0$ with delivery period $[t,t+1]$. Model (\ref{Opt4}) can be further extended to account for intraday, forward \textit{and} spot transactions, as well.
\end{rem}

\begin{rem}
The model proposed is intrinsically \textit{balance-energy neutral} for the balance group which the hydro infrastructure belongs to. A \textit{balance group} is a set of electricity meters measuring $15$ min consumption and production for net users. The \textit{transmission system operator} makes sure that every balance group is in an equilibrium state, by adding or subtracting electric energy in such a way that the total sum of energies vanishes for every quarter of an hour. Of course this comes at a certain expensive price with which the TSO charges the balance group owner, which can be (but not necessarily is) the hydro infrastructure owner as well. Thus, there is an incentive not to generate or at least to reduce balance energy, in order to minimize costs.
\end{rem}

\begin{rem}
If we assume that the utility function $U:\mathbf{R}\hookrightarrow \mathbf{R}$ can be written as as
\begin{equation}
U=r-\frac{w}{2}\rho,
\end{equation}
\noindent where $r$ is an increasing concave function, $\rho$ is an increasing convex
function and $w>0$ the risk aversion parameter, then
the optimization problems analyzed so far can be rewritten in terms of risk-reward optimization, as it is customary in financial portfolio theory.
\begin{defi}[\textbf{Risk and Reward}]
The functional
\begin{itemize}
\item $\text{Reward}:L^2(\Omega,\mathcal{A},P):\rightarrow \mathbf{R}, R\mapsto \text{Reward}(R):=\mathbb{E}_0[r(R)]$ is termed as \textbf{reward measure},
\item $\text{Risk}:L^2(\Omega,\mathcal{A},P):\rightarrow \mathbf{R}, R\mapsto \text{Risk}(R):=\mathbb{E}_0[\rho(R)]$ is termed as \textbf{risk measure}.
\end{itemize}
\end{defi}
\noindent The optimization problem (\ref{Opt2}) reads then as a trade-off between total risk and total reward
\begin{equation}\label{Opt5}
\max_{\substack{(u_t)_{t=0,\dots,T-1}\\ \text{Restrictions}}}
\;\left[\sum_{t=1}^{T}\beta_t\text{Reward}(u_{t-1}X_t)\right]-\frac{w}{2}\left[\sum_{t=1}^{T}\beta_t\text{Risk}(u_{t-1}X_t)\right].
\end{equation}
\end{rem}


\section{A Numerical Example}

\begin{figure}[!]
\centering
\includegraphics[width=10cm]{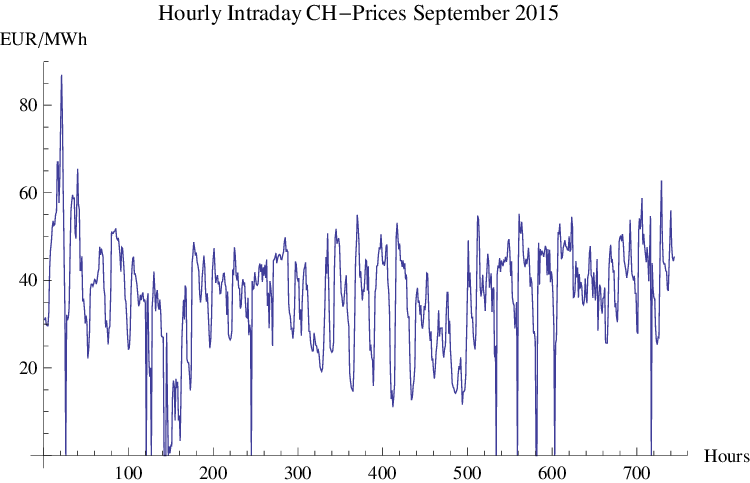}\newline
\caption{Intraday Prices}
\label{PlotIntraday}
\end{figure}

We utilize the weighted averaged September 2015 data from Epex Spot Intraday
Continuous for the CH Market, downloaded from \textit{%
www.epexspot.com/en/market-data/intradaycontinuous/intraday-table/2015-09-30/CH%
}. These weighted averaged intraday are plotted in Figure \ref{PlotIntraday}
and have descriptive statistics as in Table \ref{Stat}.\par

\begin{table}[!]
\begin{tabular}{|l|l|}
\hline
\textbf{Quantity} & \textbf{Value} \\ \hline\hline
Max & $86.91$ EUR/MWh \\ \hline
Min & $0.56$ EUR/MWh \\ \hline
Mean & $37.40$ EUR/MWh \\ \hline
Volatility & $123.87$ EUR/MWh \\ \hline
Volatility of hourly log returns & $19.97\%$ \\ \hline
\end{tabular}
\caption{Intraday prices statistics}
\label{Stat}
\end{table}

We construct a simple hydro infrastructure as described in Table \ref{Basin} and test two possible intraday price dynamics as shown in Table \ref{Intraday_Dynamics}.\par

\begin{table}[!]
\begin{tabular}{|l|l|l|}
\hline
\textbf{Symbol}&\textbf{Description} & \textbf{Value} \\ \hline\hline
$B$ & Number of basins & 1 \\ \hline
$Y^{\text{Max}}$ & Basin level maximal capacity & $160\,\text{GWh}$ \\ \hline
$Y^{\text{Min}}$ & Basin level minimal capacity & $40\,\text{GWh}$ \\ \hline
$u^{\text{Max}}$ & Turbine maximal capacity & $500\,\text{MW}$\\ \hline
$u^{\text{Min}}$ & Turbine minimal capacity (no pumping) & $0\,\text{MW}$\\ \hline
$Y_0$ & Two possible basin starting level & $80\,\text{GWh}$ and $41.5\,\text{MWh}$ \\ \hline
$i_t$ & (No) Inflow & $0\,\text{MWh}$ \\ \hline
\end{tabular}
\caption{Basin Parametrization}
\label{Basin}
\end{table}

\begin{table}[!]
\begin{tabular}{|l|l|l|}
\hline
\textbf{Model}&\textbf{Description} & \textbf{Parameters} \\ \hline\hline
Model 1 & Driftless geometric Brownian motion as in (\ref{GBM}) & $\sigma_t=19.97\%$ \\
& & $K=1$ \\ \hline
Model 2 & Driftless spread to spot as in (\ref{STS}) & $\eta_t=4.12$ EUR/MWh,\\
& & $\mathbb{E}_0[S_t]=$Sep 2015 means \\ \hline
\end{tabular}
\caption{Intraday Dynamics $X_t$ Models}
\label{Intraday_Dynamics}
\end{table}

For every day in the sample we compute the optimal dynamic strategy and the
water values for the day ahead market using the weighted average electricity
price between 08:00 and 09:00 for the current day. More precisely, we make
the choices for the lattice specified in Table \ref{Lattice_Parametrization}.

\begin{table}[!]
\begin{tabular}{|l|l|l|}
\hline
\textbf{Symbol}&\textbf{Description} & \textbf{Value} \\ \hline\hline
$t$ &Valuation time & $0$ (08:00) \\ \hline
$X_0$ &Intraday price starting value & weighted average\\
& &price 08:00-09:00\\ \hline
$t_0$ &Initial time day ahead &$24$ \\ \hline
$T$ &Final time day ahead &$48$ \\ \hline
$k$ &Number of branches out of a leaf in the lattice &$15$\\ \hline
\end{tabular}
\caption{Lattice Parametrization}
\label{Lattice_Parametrization}
\end{table}

\begin{table}[!]
\begin{tabular}{|l|l|l|}
\hline
\textbf{Utility Function}&\textbf{Definition} & \textbf{Parameters} \\ \hline\hline
Linear & $U(v):=v$ &  \\ \hline
Exponential & $U(v):=1-\exp(-\alpha v)$ & $\alpha>0$: Arrow-Pratt relative risk aversion \\ \hline
Logarithmic & $U(v):=\log(v)$ & \\ \hline
Hyperbolic & $U(v):=\frac{1}{\gamma}v^{\gamma}$ &$\gamma\in]0,1[$ \\ \hline
\end{tabular}
\caption{Utility Functions}
\label{Utility_Parametrization}
\end{table}

As expected, when the chosen starting level is $80\,\text{GWh}$  and thus the basin level constraints can never become binding, the optimal
strategy is the same for all utility functions in Table \ref{Utility_Parametrization} and reads
\begin{equation}
u_t^*(n)= 500\,\text{MWh}\quad \text{for all times $t$ and nodes $n$.}
\end{equation}

\begin{table}[h!]
\centering
\includegraphics[width=8.25cm, angle=-90]{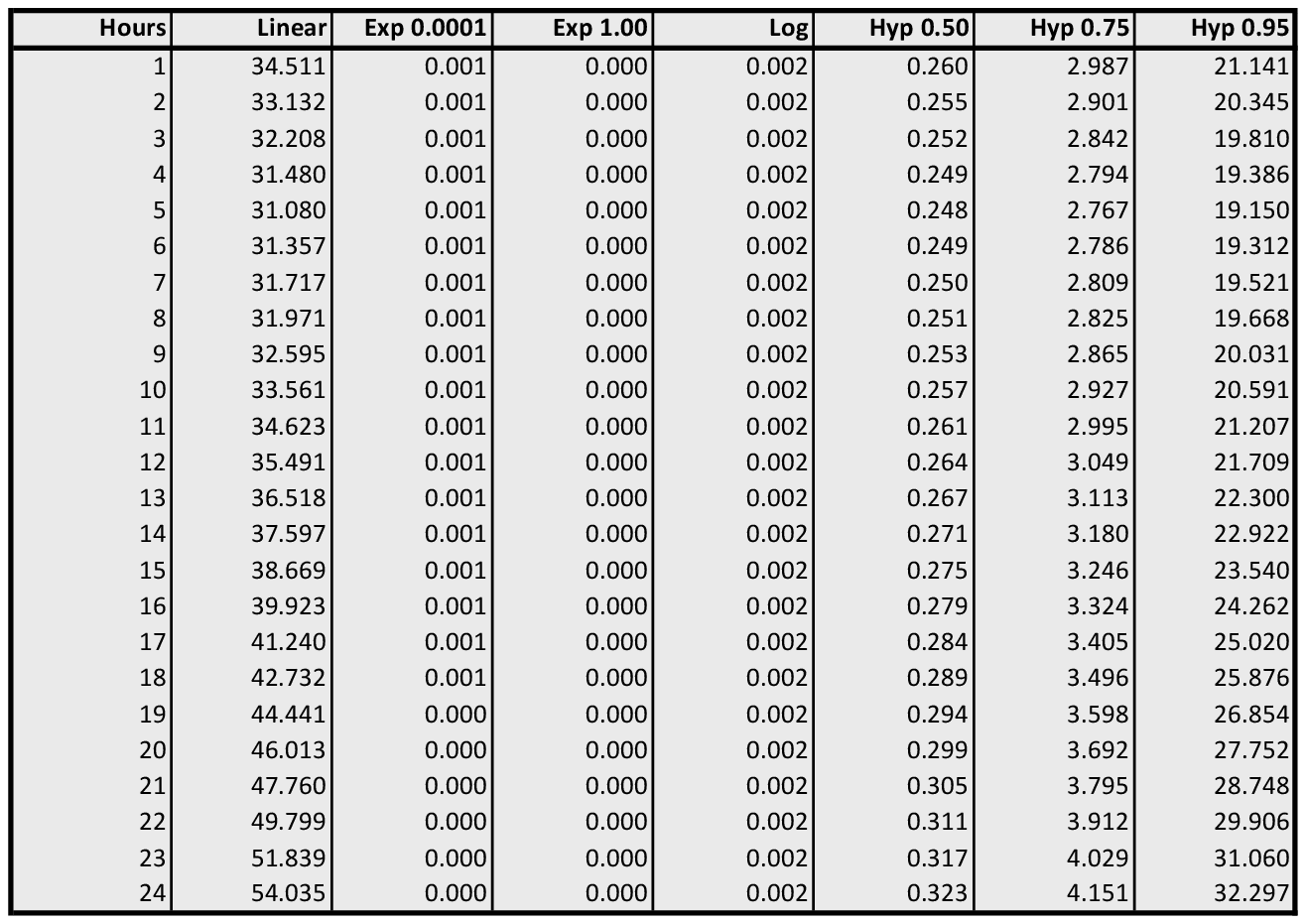}\newline
\caption{Water Values $\text{GP}_t^\text{Ask}$ for September 2, 2015,
Driftless GBM dynamics, initial basin level $80\,\text{GWh}$}
\label{WV1}
\end{table}

\begin{table}[h!]
\centering
\includegraphics[width=8.25cm, angle=-90]{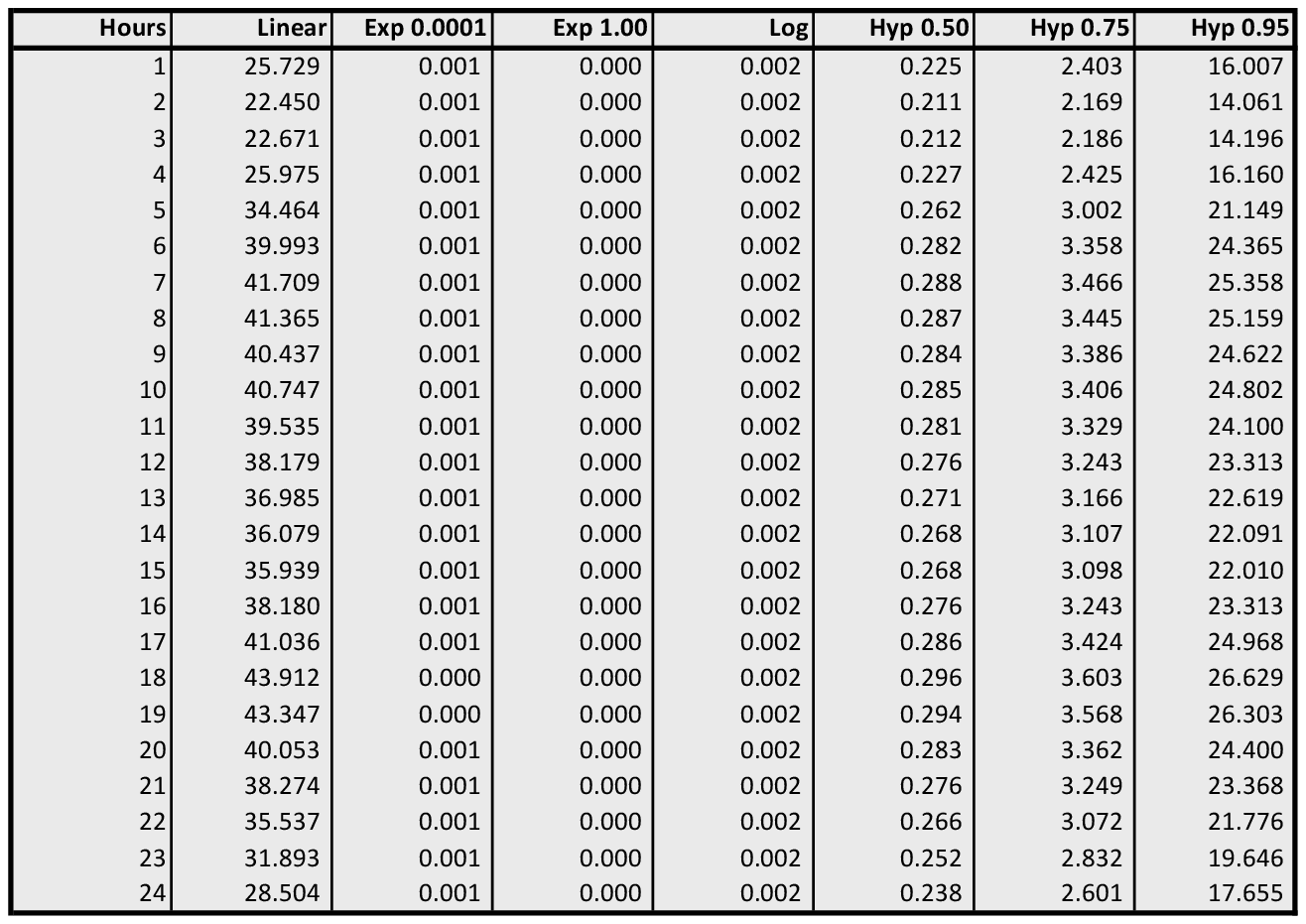}\newline
\caption{Water Values $\text{GP}_t^\text{Ask}$ for September 2, 2015, Spread
to spot Dynamics, initial basin level $80\,\text{GWh}$}
\label{WV2}
\end{table}

To back test the results for the optimal strategy we apply it to the
historical realizations of the intraday prices. More exactly, we express the
discretized optimal rules as function of the discretized intraday price of
the preceding period and compute the optimal rule with the realized price by
linear interpolation. Then, for every day in the back test, we pass through
the different hours choosing the optimal quantity of water to be turbined
according to the dynamic control rule established before. The wealth
generated for every hour for all days is depicted in Table \ref{Wealth}.
\begin{table}[h!]
\centering
\includegraphics[height=20cm]{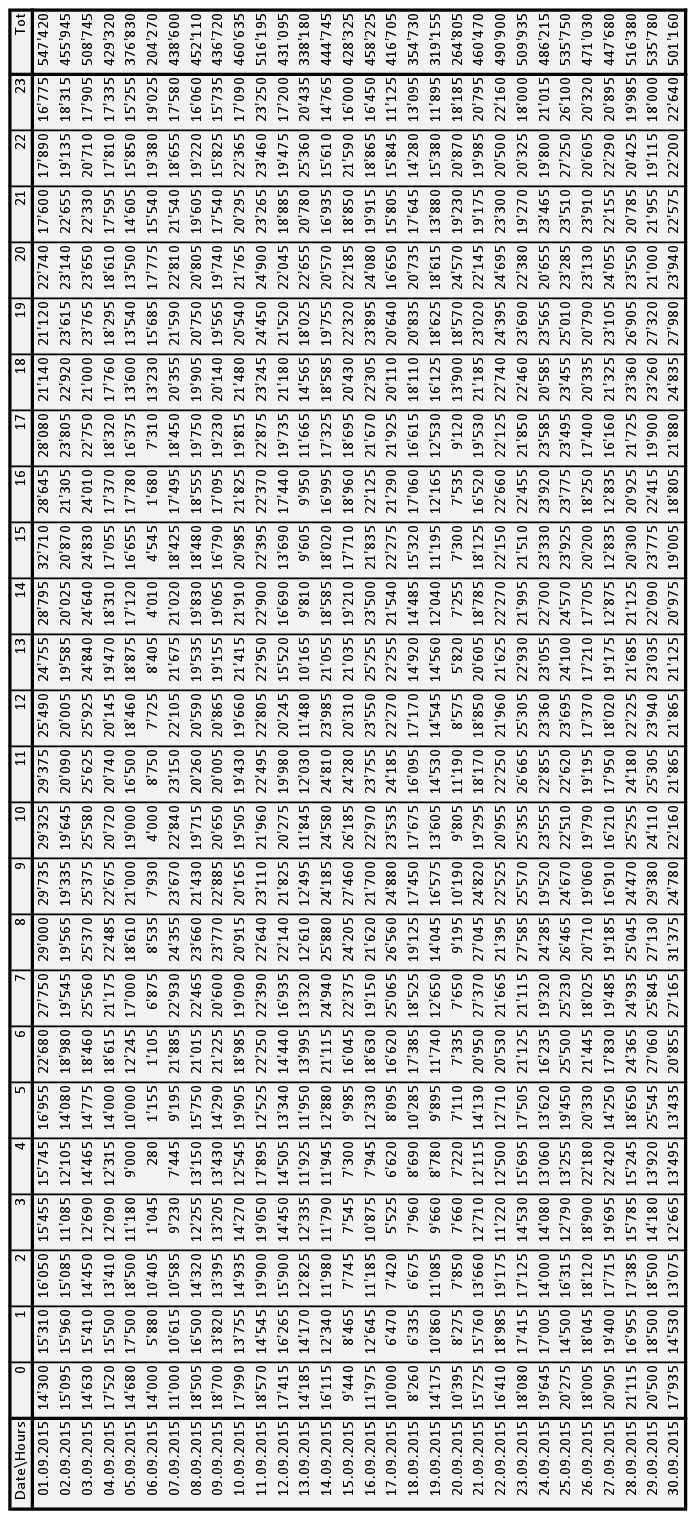}\newline
\caption{Back Test Optimal strategy, Driftless GBM dynamics, initial basin
level $80\,\text{GWh}$}
\label{Wealth}
\end{table}

\begin{figure}[h!]
\centering
\includegraphics[height=7cm, angle=-90]{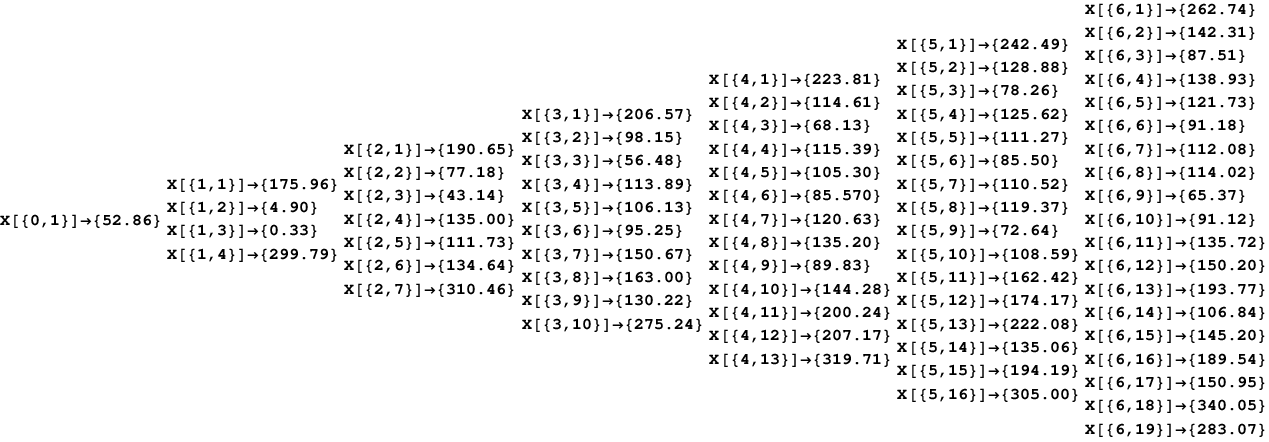}\newline
\caption{Intraday Prices, Driftless GBM dynamics}
\label{X1}
\end{figure}

\begin{figure}[h!]
\centering
\includegraphics[height=7cm, angle=-90]{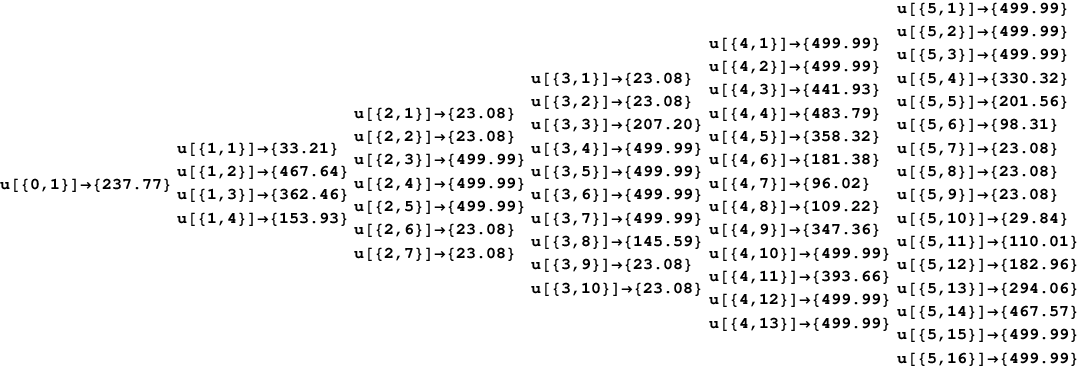}\newline
\caption{Optimal Turbined Water, Driftless GBM dynamics, initial basin level
$41.5\,\text{MWh}$}
\label{U1}
\end{figure}

\begin{figure}[h!]
\centering
\includegraphics[height=7cm, angle=-90]{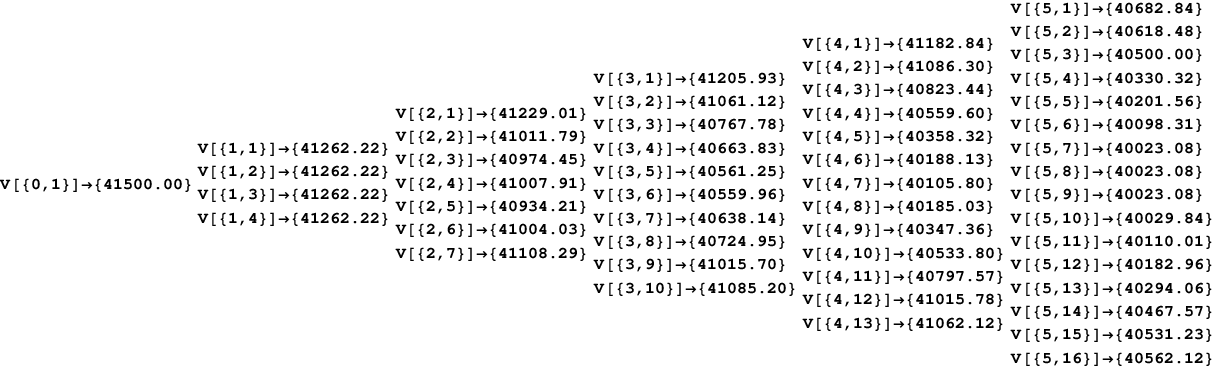}%
\newline
\caption{Basin Level, Driftless GBM dynamics, initial basin level $41.5\,%
\text{MWh}$}
\label{V1}
\end{figure}

\begin{figure}[h!]
\centering
\includegraphics[height=7cm, angle=-90]{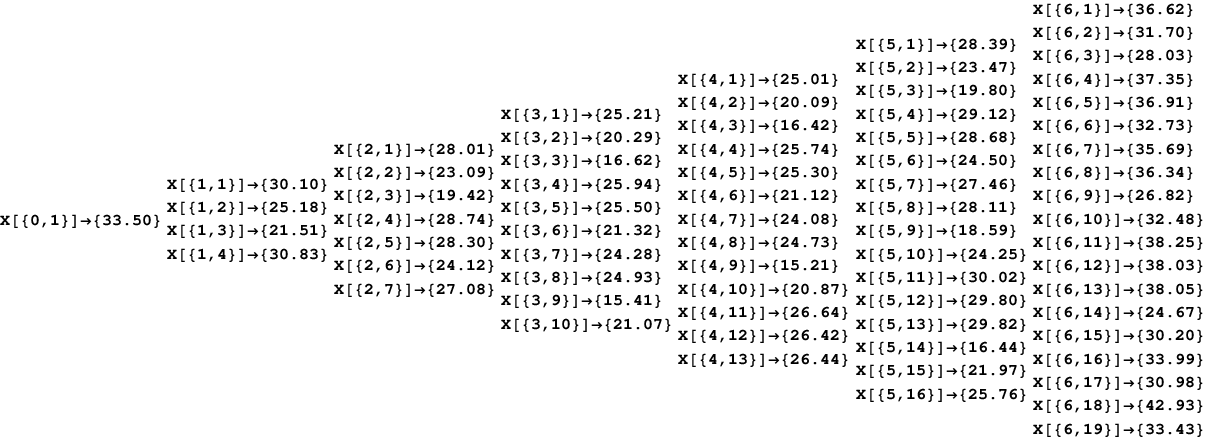}\newline
\caption{Intraday Prices, Spread To Spot dynamics }
\label{X2}
\end{figure}

\begin{figure}[h!]
\centering
\includegraphics[height=7cm, angle=-90]{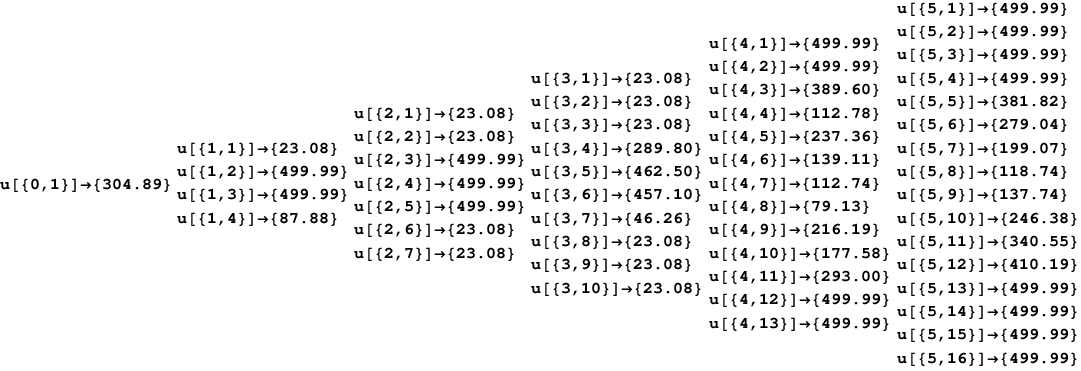}\newline
\caption{Optimal Turbined Water, Spread To Spot dynamics, initial basin
level $41.5\,\text{MWh}$}
\label{U2}
\end{figure}

\begin{figure}[h!]
\centering
\includegraphics[height=7cm, angle=-90]{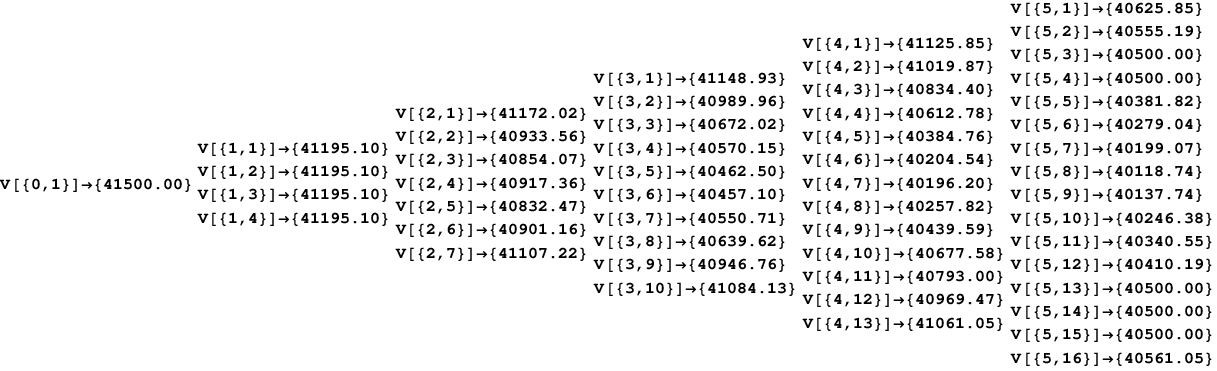}%
\newline
\caption{Basin Level, Spread To Spot dynamics, initial basin level $41.5\,%
\text{MWh}$}
\label{V2}
\end{figure}

If we set the initial basin level constraint near to the lower bound, the
optimal strategy looks different: it becomes truly stochastic, tries to
exploit the price dynamics and, of course, depends on the the utility
function chosen.
\begin{table}[!]
\begin{tabular}{|l|l|l|}
\hline
\textbf{Symbol}&\textbf{Description} & \textbf{Value} \\ \hline\hline
$t_0$ &Initial time day ahead &$0$ \\ \hline
$T$ &Final time day ahead &$6$ \\ \hline
$k$ &Number of branches out of a leaf in the lattice &$4$\\ \hline
\end{tabular}
\caption{Lattice Parametrization}
\label{Toy_Lattice_Parametrization}
\end{table}
 In the following toy examples with the parametrization specified by Table \ref{Toy_Lattice_Parametrization} we
depict the realizations of prices, optimal turbined quantities and basin
level with the hyperbolic utility function with $\gamma:=0.95$, once with
the driftless geometric Brownian motion (Figures \ref{X1}, \ref{U1}, \ref{V1}) and once with the spread to spot
dynamics (Figures \ref{X2}, \ref{U2}, \ref{V2}). In both cases we notice that the lower basin level bound becomes binding on some nodes on the final time layer
$t=T$, which -due to the intertemporal nature of the optimization- has consequences on \textit{all} earlier turbined quantities for $t=0,\dots,T-1$ in some nodes, which do not reach their possible maximum even though there is still enough water in the basin. This phenomenon is the current ``price'' of future constraints.

\begin{rem}[\textbf{Algorithm Parameter Choices}] Following Blomvall and Lindberg we choose $\mu_j:= \mu_0 \exp(-j)$ for $mu_0:=10^{-12}$. Note that we take only one Newton step before reducing $\mu_j$. As soon as $\mu_j<\mu_{\text{CP}}:=10^{-16}$ we assume that we have reached the close proximity to the so called central path and continue with Newton steps up to a maximum of $100$.
\end{rem}

\begin{rem}[\textbf{Computational time of the Mathematica prototype}] We run the prototype on a Lenovo computer with Intel Core  $i7-3740$QM CPU  $@2.70$ GHz. Typically, it takes between $4$ and $6$ minutes to compute the pedagogical examples of Figures  \ref{X1}, \ref{U1}, \ref{V1} and \ref{X2}, \ref{U2}, \ref{V2} for the toy lattice parametrization specified by Table \ref{Toy_Lattice_Parametrization}, and between $7$ and $8$ hours to compute the realistic example for the lattice parametrization specified by Table \ref{Lattice_Parametrization}. We observe that, the more constraints are binding, the longer the computational time is. Since our Mathematica code is not optimized, we are confident that a reimplementation in a faster language (e.g. C) and the utilization of better hardware can drastically improve the performance.
\end{rem}

\section{Conclusion and Further Research}

A stochastic multiperiod portfolio optimization problem in discrete time for
a generic utility function is discretized in the space dimensions by means
of a lattice. Inequality constraints are packed into the objective function
by means of a logarithmic penalty and the utility function is approximated
by its second order Taylor polynomial. A sequence of solutions of
the approximated problem converging to the optimal solution of the original
problem is constructed and coded in an algorithm in Mathematica.
We implement the algorithm on a lattice and apply it to intraday electricity trading.
We obtain:
\begin{itemize}
\item a \textit{novel, computationally efficient} implementation of a risk averse intertemporal portfolio optimization for the intraday market, and
\item \textit{deterministic water values} of an hydro infrastructure for the day ahead market bids as certainty equivalents of optimal stochastic Lagrangian multipliers corresponding to the basin level equations.
\end{itemize}
 In a next work we will:

\begin{itemize}
\item compare the lattice implementation with the grid implementation, for
both the semi-closed formula and the generic case.

\item investigate the specific case of a quadratic utility function which
needs no Newton-Scheme, being its second order Taylor polynomial the utility
function itself, and, in particular, the dynamic mean variance case, for
which in \cite{FV08} a semi-closed solution was already provided.

\item analyze the case of the maximization of the expected utility of the
cumulated values over the different time subperiods, when the utility
function is a trade-off between expectation and a dynamic risk measure, thus
allowing for Bellman's recursive approach.

\item construct an example where the pumping mode will occur in the optimal solution.

\item analyze the present costs of future constraints.

\item utilize the algorithm to compute opportunity costs to price ancillary services.
\end{itemize}

\section*{Acknowledgments}

We would like to thank Sai Anand and R\'{e}mi Janner for their hints and
their very valuable feedbacks.

\end{document}